\documentclass{article}
\usepackage{multirow}
\usepackage{booktabs}
\usepackage{amsmath,booktabs,tabularx,makecell,array}
\usepackage[utf8]{inputenc}
\usepackage{indentfirst}
\usepackage{epsfig}
\usepackage{amsmath}
\numberwithin{equation}{section}
\usepackage{subcaption}
\usepackage{authblk}
\usepackage{siunitx}
\usepackage{epsf}
\usepackage{graphicx,xcolor,overpic,mathtools}

\usepackage{amssymb}
\usepackage{multirow,lscape,array}
\usepackage{fullpage}
\usepackage{cite}
\usepackage{float}
\usepackage{enumitem}
\usepackage{amsfonts}
\usepackage{graphicx}
\usepackage{array}
\usepackage{caption}
\newcolumntype{P}[1]{>{\centering\arraybackslash}p{#1}}
\usepackage[colorlinks,linkcolor=blue,anchorcolor=red,citecolor=blue]{hyperref}
\usepackage{cite} 
\title{Numerical study of hypershadows in higher-dimensional black holes}
\author{Jianzhi Yang\footnote{jianzhi@ua.pt}}
\affil{\small \textit{Departamento de Matem\'atica da Universidade de Aveiro and} \\ 
\textit{Centre for Research and
Development in Mathematics and Applications (CIDMA),} \\ \textit{Campus de Santiago, 3810-193
Aveiro, Portugal}}

\begin{document}

\maketitle

\begin{abstract}
	We develop a fully numerical framework to compute and visualize the \emph{hypershadow}\cite{Novo:2024wyn}, the three-dimensional generalization of the black hole shadow in five-dimensional spacetimes. 
	Our method is based on backward ray tracing and allows flexible control over observer position, enabling the reconstruction of the full shadow volume. For visualization, we combine discrete sampling with surface contouring and introduce reflection difference maps on central slices to quantify mirror symmetries.
	
	Applying this method to the Schwarzschild-Tangherlini and Myers-Perry geometries, we validate the former's spherical symmetry and systematically discuss the hypershadow's dependence on observer position and black hole spin parameters. 
	We also provide compact quantitative measures for size reduction and global displacement, revealing clear monotonic trends. 
	The framework is readily extendible to other metrics and opens the way to numerical studies of more exotic objects, such as black rings and their prospective toroidal hypershadows.
\end{abstract}

\section{INTRODUCTION}
In recent years, there has been growing interest in the properties of gravity beyond four dimensions, driven largely by developments in string theory, the AdS/CFT correspondence, and models with TeV--scale extra dimensions\cite{strominger1996microscopic,aharony2000large,cavaglia2003black,kanti2004black}. In higher-dimensional spacetimes, black hole solutions exhibit qualitatively new features: the constraint of spherical horizon topology is relaxed, and the classical uniqueness theorems of four-dimensional general relativity no longer apply.
Several higher-dimensional solutions to Einstein's equations are well known, including the Tangherlini generalizations \cite{Tangherlini:1963bw} of the Schwarzschild and Reissner-Nordström black holes for 
\( D > 4\) and the Myers-Perry extension \cite{Myers:1986un} of the Kerr solution. All of these have spherical event horizons. However, they can coexist with solutions like the black ring \cite{Emparan:2008eg,emparan2002rotating}, which possesses a toroidal \( S^1 \times S^{D-3} \) topology yet shares the same mass and angular momentum---providing a direct violation of uniqueness in vacuum gravity.

As this example illustrates, adding just one extra dimension can fundamentally alter the solution space of black holes. This motivates a closer examination of which structural and observable features of black holes persist or fail to persist in higher dimensions. One particularly intriguing aspect is the black hole shadow, whose properties in higher dimensions remain only partially understood.
The black hole shadow is the apparent image of a black hole projected onto the observer's celestial sphere. In a four-dimensional spacetime, the shadow is a two-dimensional region on the surface of the celestial sphere. 
In a five-dimensional spacetime, the black hole shadow lies on a three-dimensional celestial sphere, giving rise to a genuinely three-dimensional structure, which we refer to as the \emph{hypershadow}.

There are two common approaches to computing black hole shadows: analytical and numerical. The analytical method typically uses the Hamilton-Jacobi formalism to derive separable geodesic equations and compute the shadow contour. The numerical method relies on backward ray tracing to reconstruct the shadow shape by simulating photon trajectories from the observer's screen. While both methods have been extensively applied to two-dimensional shadows in four-dimensional spacetimes\cite{bardeen1973timelike,cunha2018shadows,Papnoi:2014aaa, Vagnozzi:2022moj}, the three-dimensional hypershadow has so far only been explored analytically, most notably in Refs.~\cite{Novo:2024wyn,estrada2025five}. A fully numerical approach to constructing hypershadows remains undeveloped. In this work, we introduce such a numerical method to compute the shape of the hypershadow in higher-dimensional spacetimes.

With the numerical method developed here, we are able to compute hypershadows for a broader class of black hole spacetimes. This allows us to systematically study how hypershadows deform in response to different observer position, such as changes in orientation, shape, and position.

To enhance visual interpretation, we propose an improved visualization scheme that combines discrete point sampling with surface contouring, offering clearer representations of the three-dimensional structure of hypershadows. Furthermore, by analyzing the reflection difference of their central slices, we introduce a quantitative measure of mirror symmetry in hypershadows. We then discuss the effects of observer position on the black hole shadow.
These methodological advances lay the groundwork for future investigations into more exotic black hole geometries. In particular, they open the possibility of numerically exploring whether the hypershadow of a black ring exhibits a topological torus structure.

A key novelty of this work lies in the systematic analysis of the observer inclination angle $\theta_{\rm o}$ in higher dimensions. 
In five-dimensional Myers-Perry black holes we introduce quantitative measures of hypershadow deformation: a distortion parameter $\delta_s$, capturing size reduction relative to the Schwarzschild-Tangherlini baseline, and a displacement parameter $\eta$, relevant in the singly rotating case. 
We find that $\delta_s$ grows with the spin and, in the single-rotation geometry, also with inclination, while $\eta$ quantifies a symmetry-breaking shift absent in the cohomogeneity-one case. 
Together, these parameters show how $\theta_{\rm o}$ influences the rotational structure of higher-dimensional black holes.

This paper is organized as follows. In Sec.~\ref{sec2}, we introduce the backward ray-tracing method used to compute the hypershadow. 
In Sec.~\ref{sec3}, we present the numerical hypershadows of several higher--dimensional black hole solutions.  In Sec.~\ref{effects}, we study the effects of observer position on the black hole shadow. In Sec.~\ref{sec5}, we present our conclusions.

\section{BACKWARD RAY TRACING}
\label{sec2}
\subsection{Local observer basis}
The notation for the coordinate set $\chi = \{t, x, \theta, \phi, \psi\}$ was chosen in order to appear spherical-like. The radial coordinate $x$ is the square of the usual spherical coordinate $r$, i.e. $x=r^2$. This will allow us to simplify calculations and make many of the expressions more compact. The time coordinate $t$ and the azimuthal coordinates $\phi$ and $\psi$ adapt to the corresponding Killing vector, namely, \( \partial_t, \partial_\phi \), and \( \partial_\psi \). There is a conserved quantity of motion along geodesics associated with each Killing vector, 
\( p_t = -E \), \( p_\phi = \Phi \), and \( p_\psi = \Psi \), respectively.

In this work, we adopt a particular orthonormal basis associated with a reference frame that has zero axial angular momentum with respect to spatial infinity. 
Such a frame is commonly referred to as the zero angular momentum observer(ZAMO) frame\cite{Bardeen:1973tla}, and is frequently used in black hole spacetimes to describe locally nonrotating observers.
The ZAMO frame defines a natural decomposition of the observer's orthonormal basis vectors in terms of the coordinate basis vectors \( \{ \partial_t, \partial_x, \partial_\theta, \partial_\phi, \partial_\psi \} \). 
While this decomposition is not unique, we adopt the following specific form in this paper:
\begin{subequations} 
\begin{align}
\hat{e}_{(t)} &= A^t \partial_t + B^t \partial_\phi + C^t \partial_\psi , \\
\hat{e}_{(\phi)} &= A^\phi \partial_\phi + B^\phi \partial_\psi , \\
\hat{e}_{(\psi)} &= A^\psi \partial_\psi , \\
\hat{e}_{(x)} &= A^x \partial_x , \\
\hat{e}_{(\theta)} &= A^\theta \partial_\theta .
\end{align}
\end{subequations}
Since the observer perceives the spacetime locally as Minkowski, the basis vectors are normalized according to flat space conditions
\begin{equation}
\hat{e}_{(\mu)} \cdot \hat{e}_{(\nu)} = \hat{e}^{\alpha}_{(\mu)} g_{\alpha\beta} \hat{e}^{\beta}_{(\nu)} = \eta_{(\mu)(\nu)} ,
\end{equation}
where $\eta$ denotes the Minkowski metric. \\
It is always possible to choose the non-Killing coordinates to be orthogonal so the corresponding coefficients would read
\begin{equation}
    A^x=\frac{1}{\sqrt{g_{xx}}}, \quad A^\theta=\frac{1}{\sqrt{g_{\theta\theta}}}.
\end{equation}
The coefficient \( A^\psi \) is defined as
\begin{equation}
A^\psi=\frac{1}{\sqrt{g_{\psi\psi}}},
\label{A_psi}
\end{equation}
and the remaining coefficients can be determined by imposing Minkowski normalization together with the specific form of the metric.\\
The locally measured momenta are
\begin{subequations}
\begin{align}
p^{(t)} &= -\hat{e}^\mu_{(t)} p_\mu = A^t E - B^t \Phi - C^t \Psi , \\
p^{(\phi)} &= \hat{e}^\mu_{(\phi)} p_\mu = A^\phi \Phi + B^\phi \Psi , \\
p^{(\psi)} &= \hat{e}^\mu_{(\psi)} p_\mu = \frac{\Psi}{\sqrt{g_{\psi\psi}}}, \\
p^{(x)} &= \hat{e}^\mu_{(x)} p_\mu = \frac{p_x}{\sqrt{g_{xx}}} , \\
p^{(\theta)} &= \hat{e}^\mu_{(\theta)} p_\mu = \frac{p_\theta}{\sqrt{g_{\theta\theta}}} .
\end{align}
\end{subequations}

\subsection{Impact parameters}
The shadow of a four-dimensional black hole is typically represented as a two-dimensional shape on an image plane with Cartesian coordinates \( \{X, Y\} \), which correspond to two observation angles \( \{\alpha, \beta\} \). In the five-dimensional case, this generalizes to a three-dimensional hypershadow, represented as a volume in Cartesian coordinates \( \{X, Y, Z\} \), where the additional coordinate \( Z \) corresponds to a third observation angle \( \zeta \).

We, therefore, construct a 3D image space using Cartesian coordinates \( \{X, Y, Z\} \), which represent the impact parameters of light rays and are proportional to the observation angles \( \{\alpha, \beta, \zeta\} \). The observer is assumed to be located at coordinates \( (x_{\rm o}, \theta_{\rm o}) \), positioned very far from the black hole, i.e., \( x_{\rm o} \gg \mu \), where \( \mu \) denotes the mass scale of the black hole. Throughout the paper we refer to $x_{\rm o}$ as the observer radial coordinate and to $\theta_{\rm o}$ as the observer inclination angle; their influence on the hypershadow will be analyzed in Sec.~\ref{effects}.

In the far-away observation limit \( x_{\rm o} \to \infty \), the observation angles decay as \( \sim 1/r_{\rm o} = 1/\sqrt{x_{\rm o}} \), following the same scaling law as in the four-dimensional case. Accordingly, the impact parameters are defined as
\begin{equation}
X = -\sqrt{x_{\rm o}} \, \beta , \quad
Y = -\sqrt{x_{\rm o}} \, \zeta , \quad
Z = \sqrt{x_{\rm o}} \, \alpha .
\end{equation}

The components of the photon's physical momentum satisfy the constraint
$\left[p^{(t)}\right]^2 = \mathbf{p}^2,$
where \( \mathbf{p} \) denotes the norm of the spatial momentum. Explicitly, this norm is given by:
\begin{equation}
\mathbf{p}^2 = \sum \left[p^{(A)}\right]^2, \quad A = \{x, \theta, \phi, \psi\}.
\label{p_decomposition}
\end{equation}

Therefore, the spatial components of the photon's momentum can be parametrized in terms of the observation angles as follows:
\begin{subequations}\label{eqmomentum}
\begin{align}
p^{(x)} &= \mathbf{p} \cos\alpha \cos\beta \cos\zeta, \\
p^{(\theta)} &= \mathbf{p} \sin\alpha, \\
p^{(\phi)} &= \mathbf{p} \cos\alpha \sin\beta,  \\
p^{(\psi)} &= \mathbf{p} \cos\alpha \cos\beta \sin\zeta. 
\end{align}
\end{subequations}
Here, the origin of the observation angles, \( \{\alpha, \beta, \zeta\} = \{0, 0, 0\} \), corresponds to a direction pointing radially away from the black hole. In the far--distance limit, the observation angles become very small, and the spatial momentum components can be approximated as
\begin{equation}
p^{(\theta)} \simeq \mathbf{p} \, \alpha , \quad
p^{(\phi)} \simeq \mathbf{p} \, \beta , \quad
p^{(\psi)} \simeq \mathbf{p} \, \zeta .
\end{equation}

In natural units, the local energy of the photon is measured as \( \varepsilon = \mathbf{p} \), which is also equal to the time component of the four-momentum, i.e. ,\ \( \varepsilon = p^{(t)} \). Therefore, the impact parameters can be written as
\begin{equation}
X = -\sqrt{x_{\rm o}} \, \frac{p^{(\phi)}}{p^{(t)}} , \quad
Y = -\sqrt{x_{\rm o}} \, \frac{p^{(\psi)}}{p^{(t)}} , \quad
Z = \sqrt{x_{\rm o}} \, \frac{p^{(\theta)}}{p^{(t)}}.
\end{equation}

\subsection{Geodesic equation and initial condition}
The motion of null geodesics can be described by Hamilton's equations:
\begin{equation}
   \dot{\chi}^\mu = \frac{\partial \mathcal{H}}{\partial p_\mu}, \quad \dot{p}_\mu = -\frac{\partial \mathcal{H}}{\partial \chi ^ \mu},
\end{equation}
where the Hamiltonian is given by
\begin{equation}
    \mathcal{H} = \frac{1}{2} g^{\mu\nu} p_\mu p_\nu = 0.
\end{equation}
Since \( p_t \), \( p_\phi \), and \( p_\psi \) are conserved quantities, they can be identified with \( -E \), \( \Phi \), and \( \Psi \), respectively. This leads to a system of seven first-order differential equations in terms of the canonical variables \( \{\chi^\mu, p_\mu\} \). Such a formulation is particularly suitable for numerical geodesic integration, which determines whether a given light ray trajectory intersects the black hole.

We next specify the initial conditions required for geodesic equations. The initial values for the coordinates \( x(0) \) and \( \theta(0) \) are chosen according to the observer position \( (x_{\rm o}, \theta_{\rm o}) \), while the remaining coordinates \( t(0) \), \( \phi(0) \), and \( \psi(0) \) can be set to zero without loss of generality due to the corresponding conserved momenta.

Our focus is thus, on determining the initial conditions for the nonconserved momentum components \( p_x(0) \) and \( p_\theta(0) \). By combining the five-momentum projections Eqs.~\eqref{eqmomentum} with the spatial momentum norm given in Eq.~\eqref{p_decomposition}, we obtain

\begin{subequations}
\begin{align}
    p_x &= \mathbf{p} \sqrt{g_{xx}} \cos\alpha \cos\beta \cos\zeta,  \\
    p_\theta &= \mathbf{p} \sqrt{g_{\theta\theta}} \sin\alpha, \\
    \Psi &= \mathbf{p} \sqrt{g_{\psi\psi}} \cos\alpha \cos\beta \sin\zeta, \\
    \Phi &= \frac{p^{(\phi)}}{A^\phi} - \frac{B^\phi}{A^\phi A^\psi} p^{(\psi)}\notag \\
         &= \frac{\mathbf{p} \cos\alpha \sin\beta}{A^\phi} - \frac{B^\phi}{A^\phi A^\psi} \mathbf{p} \cos\alpha \cos\beta \sin\zeta, \\
    E &= \frac{p^{(t)}}{A^t} + \frac{B^t}{A^t} \left( \frac{p^{(\phi)}}{A^\phi} - \frac{B^\phi}{A^\phi A^\psi} p^{(\psi)} \right) + \frac{C^t}{A^t A^\psi} p^{(\psi)}\notag \\
      &= \frac{\mathbf{p}}{A^t} + \frac{B^t}{A^t} \left( \frac{\mathbf{p} \cos\alpha \sin\beta}{A^\phi} - \frac{B^\phi}{A^\phi A^\psi} \mathbf{p} \cos\alpha \cos\beta \sin\zeta \right) + \frac{C^t}{A^t A^\psi} \mathbf{p} \cos\alpha \cos\beta \sin\zeta.
\end{align}
\end{subequations}

Given the initial coordinates \( (x_{\rm o}, \theta_{\rm o}) \) and a chosen set of observation angles \( \{\alpha, \beta, \zeta\} \), we compute the conserved quantities \( \{E, \Phi, \Psi\} \) and the corresponding momentum components \( \{p_x(0), p_{\theta}(0)\} \). These quantities, together with the observer position, specify the initial conditions for the geodesic system. We numerically integrate the resulting first-order differential equations using \textit{Mathematica}'s built-in \textit{NDSolve} function, solving for the coordinates \( x^\mu(\lambda) \) as functions of the affine parameter \( \lambda \).

The integration proceeds forward in \( \lambda \), and is terminated either when the radial coordinate \( x(\lambda) \) reaches the event horizon (signaling a capture by the black hole), or when it exceeds a large cutoff value (indicating escape to infinity). We employ adaptive step-size control to ensure numerical stability and precision throughout the integration. This setup allows us to systematically track the fate of each light ray emitted by the observer and determine the black hole shadow as seen from the given observation point.

\subsection{3D hypershadow visualization with 2D shadow comparison}
In traditional two-dimensional black hole shadow visualizations, such as those in Kerr spacetime\cite{Cunha:2016bpi}, the image is constructed by dividing the observation plane into small square pixels, each colored according to whether the corresponding light ray falls into the black hole. By increasing the resolution (i.e. using more, smaller pixels), one can obtain a sharp and detailed shadow image. This method works well in 2D, where the shadow is confined to a plane.
\begin{figure}[H]
  \centering
  \begin{subfigure}[t]{0.40\textwidth}
    \centering
    \includegraphics[width=0.7\textwidth]{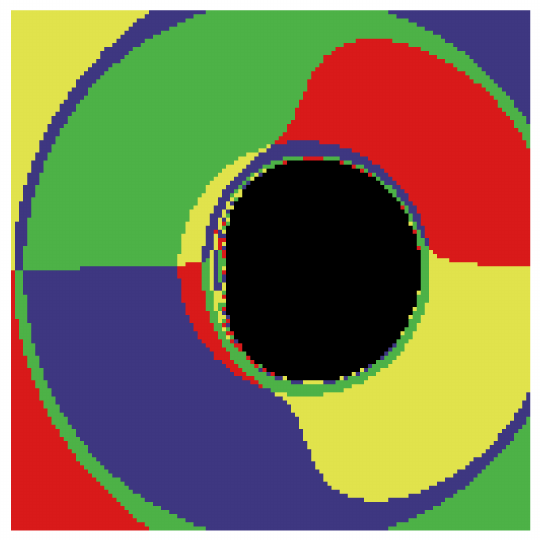}
    \caption{$N_{pix}=128$}
    \label{fig:kerr1}
  \end{subfigure}
    \hspace{0\textwidth}
  \begin{subfigure}[t]{0.40\textwidth}
    \centering
    \includegraphics[width=0.7\textwidth]{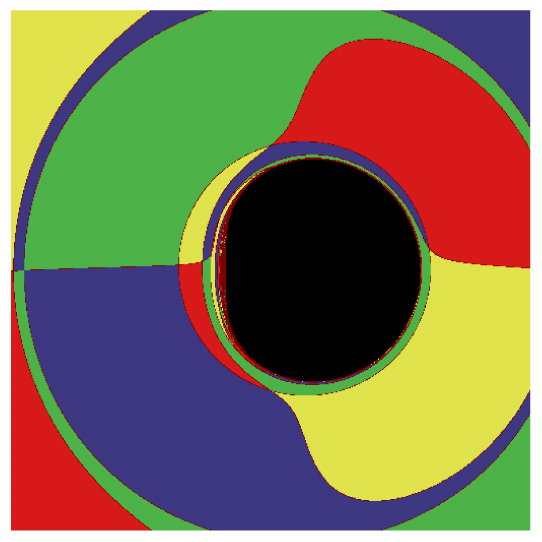}
    \caption{$N_{pix}=1024$}
    \label{fig:kerr2}
  \end{subfigure}
  \caption{ Two-dimensional Kerr black hole shadows visualized with different pixel resolutions. \\
}
  \label{fig:kerr_combined}
\end{figure}
Fig.~(\ref{fig:kerr1}) uses a lower resolution, while Fig.~(\ref{fig:kerr2}) employs a higher resolution, resulting in a sharper and more refined shadow boundary. Here, \( N_{\text{pix}} \) denotes the number of pixels along each spatial axis of the image grid. This illustrates the effectiveness of traditional 2D shadow visualization techniques when the shadow is confined to a plane.

However, when this approach is extended to visualize three-dimensional hypershadows in higher-dimensional spacetimes, the pixel-based method becomes inadequate. Even if we assign transparent colors to points where light rays escape and reserve opaque (e.g., black) colors for infalling ones, the result is still a solid, opaque volume of black voxels. This obscures the structure of the hypershadow, making it impossible to intuitively perceive its geometric shape in three dimensions.

To overcome this limitation, we propose a point-based visualization method. Instead of coloring volumetric pixels, we represent each sampling point as a discrete dot. The black dots correspond to light rays that fall into the black hole, while transparent points indicate escape. To enhance visibility of the hypershadow's shape, we further connect the boundary of the black region using blue line segments. This approach reveals not only the outer boundary, but also the overall spatial structure of the hypershadow, offering a clearer and more informative visualization than traditional voxel-based rendering.

\section{HYPERSHADOW}
\label{sec3}

In this section, we present the numerical hypershadows corresponding to several five-dimensional black hole solutions, including the Schwarzschild-Tangherlini and Myers-Perry geometries. For each case, we reconstruct the three-dimensional shadow using the ray-tracing method developed earlier and examine its global structure and symmetry properties. The aim is to demonstrate the effectiveness of our method and to highlight the distinct hypershadow features associated with static and rotating spacetimes.

\subsection{Schwarzschild-Tangherlini black hole}

In 1963, Tangherlini derived this static, spherically symmetric and asymptotically flat vacuum solution as a generalization of the Schwarzschild black hole~\cite{Tangherlini:1963bw}. 
Strong gravitational lensing in the Tangherlini spacetime has been analyzed in ~\cite{Tsukamoto:2014dta, Singh:2017vfr}.
We first introduce the five-dimensional Schwarzschild-Tangherlini black hole:
\begin{equation}
ds^2 = -\left(1 - \frac{\mu^2}{x}\right) dt^2 
       + \frac{1}{1 - \frac{\mu^2}{x}} dx^2
       + x\, d\theta^2 
       + x \sin^2\theta\, d\phi^2 
       + x \cos^2\theta\, d\psi^2 .
\end{equation}
The mass parameter $\mu$ is related to the black hole mass by
\begin{equation}\label{Mass}
M = \frac{3\pi \mu^{2}}{8 G}.
\end{equation}
Throughout this work, we set $\mu = 1$ for convenience,
and the black hole horizon $x_H$ is obtained by solving $g_{tt}(x_H)=0$, giving $x_H = \mu^2$.

At this point, the locally treated basic $\{\hat{e}_{(t)},\hat{e}_{(\phi)}\}$ and the corresponding measured momenta $\{ p^{(t)}, p^{(\phi)} \}$ can be expressed in a simplified form:
\begin{subequations}
\begin{align}
\hat{e}_{(t)} &= A^t \partial_t = \frac{1}{\sqrt{-g_{tt}}} \partial_t, \quad p^{(t)} = \frac{E}{\sqrt{-g_{tt}}} ,\\
\hat{e}_{(\phi)} &= A^\phi \partial_\phi =\frac{1}{\sqrt{g_{\phi\phi}}} \partial_\phi,\quad p^{(\phi)} = \frac{\Phi}{\sqrt{g_{\phi\phi}}} . 
\end{align}
\end{subequations}
With the linear momentum $\mathbf{p}$ decomposition Eq.~\eqref{p_decomposition}, the conserved quantity of motion $\{\Phi, E\}$ can be written in a simplified form:
\begin{subequations}
\begin{align}
    \Phi &= \mathbf{p} \sqrt{g_{\phi\phi}}\cos\alpha \sin\beta\\
    E &= \mathbf{p} \sqrt{-g_{tt}}
\end{align}
\end{subequations}
Using the Hamilton equations together with the initial conditions, we employ the backward ray-tracing method to obtain the 3D hypershadow.

\begin{figure}[H]
  \centering
  \begin{subfigure}[t]{0.4\textwidth}
    \centering
    \includegraphics[width=0.6\textwidth]{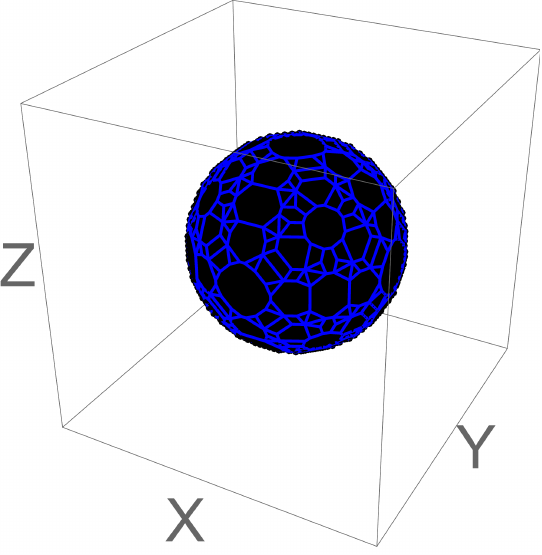}
    \caption{$N_{\rm pix}=64$}
  \end{subfigure}
  \hspace{0.00\textwidth}
  \begin{subfigure}[t]{0.4\textwidth}
    \centering
    \includegraphics[width=0.6\textwidth]{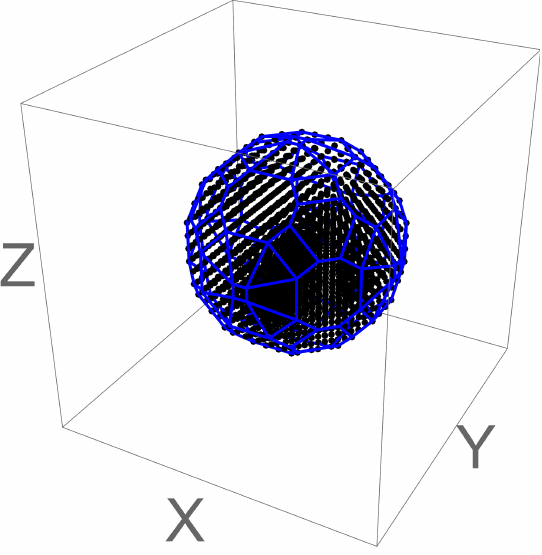}
    \caption{$N_{\rm pix}=24$}
  \end{subfigure}

  \caption{Hypershadows of the Schwarzschild-Tangherlini black hole 
 rendered with different $N_{\rm pix}$. The observer position (the initial condition) is located at $\{65,\pi/2-1/20 \}$.}
  \label{tangherlini_hypershadow}
\end{figure}

As shown in Fig.~\ref{tangherlini_hypershadow}, the Schwarzschild-Tangherlini black hole hypershadow appears perfectly spherical from the three-dimensional perspective. To rigorously confirm this spherical symmetry, we fix the observation angles to $\{\alpha, \beta, \zeta\} = 0$ and compute the cross-sectional slices through the center of the hypershadow in the $XY$, $XZ$, and $YZ$ planes.
\begin{figure}[H]
  \centering

  \begin{subfigure}[t]{0.3\textwidth}
    \centering
    \includegraphics[width=0.6\linewidth]{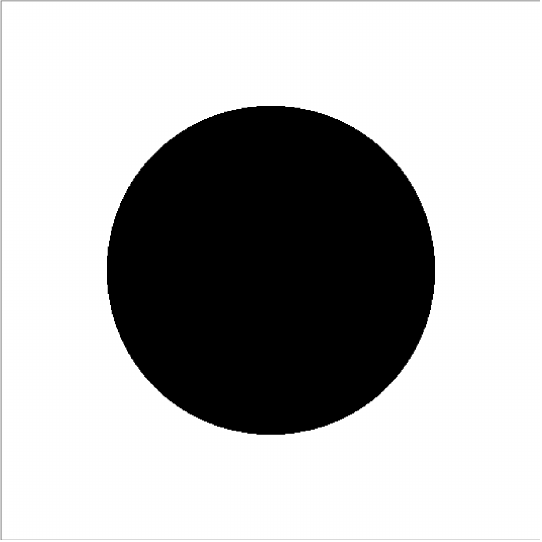}
    \caption{Central slices }
  \end{subfigure}
  \hfill
  \begin{subfigure}[t]{0.3\textwidth}
    \centering
    \includegraphics[width=0.6\linewidth]{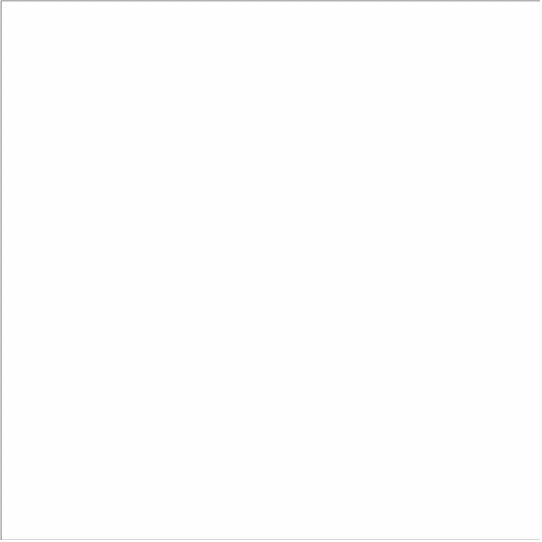}
    \caption{H-$\Delta$}
  \end{subfigure}
  \hfill
  \begin{subfigure}[t]{0.3\textwidth}
    \centering
    \includegraphics[width=0.6\linewidth]{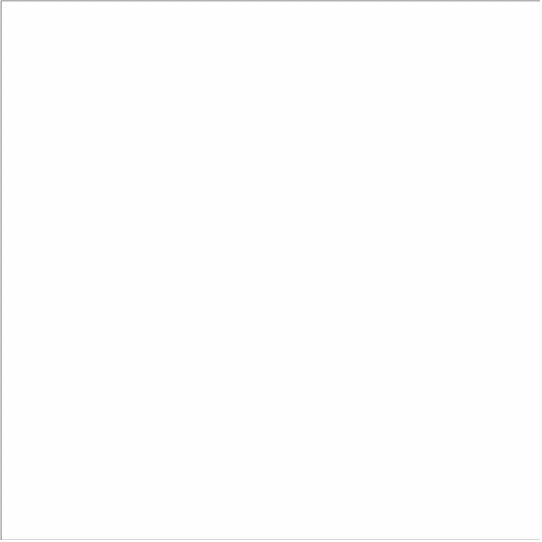}
    \caption{V-$\Delta$}
  \end{subfigure}

\caption{
(a) Central slices in the $XY$, $XZ$, and $YZ$ planes. 
(b) H-$\Delta$: difference to the left-right mirror. 
(c) V-$\Delta$: difference to the up-down mirror. 
Resolution $N_{\rm pix}=512$.}
\label{Tangherlini_slices}
\end{figure}
The symbols ``H-$\Delta$'' and ``V-$\Delta$'' denote pixelwise difference maps under mirror reflections of each 2D slice.
Specifically, for a slice plotted with the $X$ axis horizontal and $Y$ axis vertical:
\begin{itemize}
\item horizontal reflection (H-$\Delta$) is the left-right flip across the vertical midline $X\!\mapsto\! -X$, and
\item vertical reflection (V-$\Delta$) is the up-down flip across the horizontal midline $Y\!\mapsto\! -Y$.
\end{itemize}
For each slice, we reflect the image about the indicated axis and then take the absolute difference with the original to highlight departures from exact mirror symmetry. Black pixels mark locations where the reflected slice differs from the original hypershadow slice.
For the $XZ$ and $YZ$ slices, the same convention applies i.e., horizontal means reflection across the vertical axis of the displayed slice, and vertical means across the horizontal axis.

Figure.~\ref{Tangherlini_slices} displays the central slices together with H-$\Delta$ and V-$\Delta$ (definitions above). 
The vanishing of both maps (up to numerical tolerance) indicates exact reflection symmetry in each coordinate plane.

These findings not only confirm the expected spherical symmetry of the Schwarzschild-Tangherlini hypershadow, but also validate the precision and reliability of our numerical ray-tracing framework. This benchmark result serves as a foundation for exploring more intricate geometries in higher-dimensional rotating spacetimes.

\subsection{Myers-Perry black hole}
In 1986, Myers and Perry derived a stationary, axisymmetric, and asymptotically flat vacuum solution as a generalization of the Kerr black hole to higher dimensions\cite{Myers:1986un}. The 5D Myers-Perry solution is given by the line element:
\begin{equation}\label{Myers_Perry}
\begin{aligned}
ds^2 =\ & -dt^2 + (x + a^2)\sin^2\theta\, d\phi^2 + (x + b^2)\cos^2\theta\, d\psi^2 \\
& + \frac{\mu^2}{\rho^2}\left(dt + a \sin^2\theta\, d\phi + b \cos^2\theta\, d\psi \right)^2 
+ \frac{\rho^2}{4\Delta} dx^2 + \rho^2 d\theta^2,
\end{aligned}
\end{equation}
with 
\begin{equation}
\rho^2 = x + a^2 \cos^2\theta + b^2 \sin^2\theta, 
\quad 
\Delta = (x + a^2)(x + b^2) - \mu^2 x.
\end{equation}
The black hole mass is controlled by the same parameter $\mu$ introduced in Eq~(\ref{Mass}), and its two angular momenta are
\begin{equation}
J_a = \frac{2}{3} M a, \quad J_b = \frac{2}{3} M b,
\end{equation}
while $a$ and $b$ are the spin parameters associated with two independent rotation planes. These parameters are bounded by the condition $\mu \geq |a| + |b|$, which ensures the existence of an event horizon.
The angular coordinates \( \phi \) and \( \psi \) range over the interval \([0, 2\pi]\), while \( \theta \) takes values in \([0, \pi/2]\).\\
The black hole horizon is located at \( x_H = x_+ \), where
\begin{equation}
x_{\pm} = \frac{1}{2} \left[ \mu^2 - a^2 - b^2 \pm \sqrt{(\mu^2 - a^2 - b^2)^2 - 4a^2 b^2} \right].
\end{equation}
The metric (\ref{Myers_Perry}) is invariant under the following transformation:
\begin{equation}
a \leftrightarrow b, \quad \theta \leftrightarrow \left( \frac{\pi}{2} - \theta \right), \quad \phi \leftrightarrow \psi.
\end{equation}
It possesses three Killing vectors, \( \partial_t \), \( \partial_\phi \), and \( \partial_\psi \). For \( a = b \), the metric has two additional Killing vectors:
\begin{equation}
\cos \bar{\theta} \, \partial_{\bar{\theta}} - \cot \bar{\theta} \sin \bar{\phi} \, \partial_{\bar{\phi}} + \frac{\sin \bar{\phi}}{\sin \bar{\theta}} \, \partial_{\bar{\psi}}
\end{equation}
and
\begin{equation}
- \sin \bar{\phi} \, \partial_{\bar{\theta}} - \cot \bar{\theta} \cos \bar{\phi} \, \partial_{\bar{\phi}} + \frac{\cos \bar{\phi}}{\sin \bar{\theta}} \, \partial_{\bar{\psi}},
\end{equation}
where 
\begin{equation}
\bar{\phi} = \psi - \phi, \quad \bar{\psi} = \psi + \phi, \quad \bar{\theta} = 2\theta.
\end{equation}

\subsubsection{Cohomogeneity-one solution: \texorpdfstring{$a=b$}{a=b}}
The shadow of the five-dimensional Myers-Perry black hole was first investigated through two-dimensional projection images by
\cite{Papnoi:2014aaa}. Recently, the work \cite{Novo:2024wyn} provided an analytical description of the hypershadow boundary in the cohomogeneity-one case, with $a=b$. We compute the hypershadow numerically, investigate its physical features, and verify the consistency with the analytical results.

\begin{figure}[H]
  \centering
  \begin{subfigure}[t]{0.3\textwidth}
    \centering
    \includegraphics[width=0.6\textwidth]{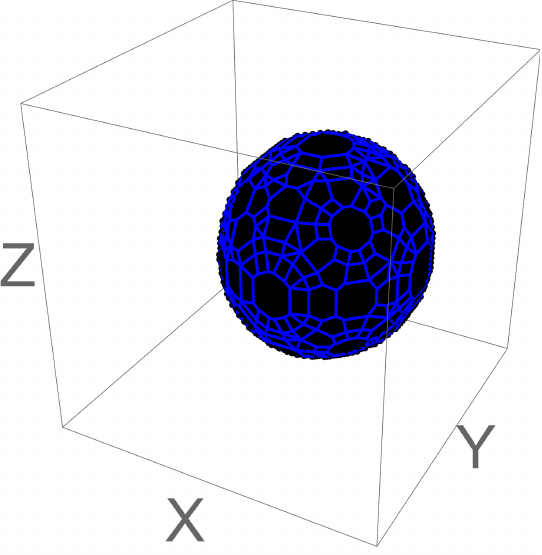}
    \caption{$N_{\rm pix}=64$}
  \end{subfigure}
  \hspace{0.00\textwidth}
  \begin{subfigure}[t]{0.3\textwidth}
    \centering
    \includegraphics[width=0.6\textwidth]{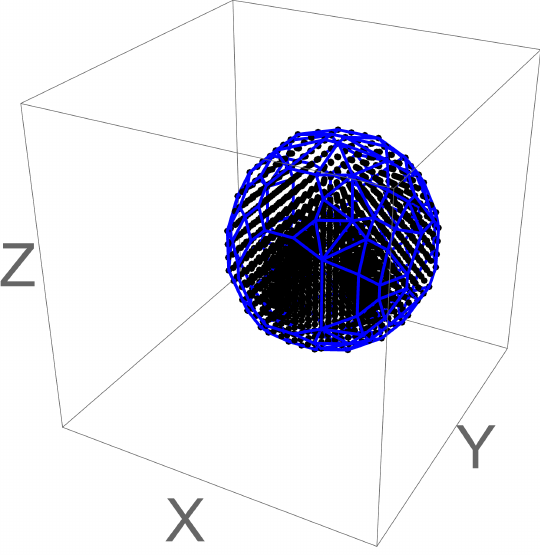}
    \caption{$N_{\rm pix}=24$}
  \end{subfigure}

  \vspace{0.5cm}

  \begin{subfigure}[t]{0.3\textwidth}
    \centering
    \includegraphics[width=0.6\textwidth]{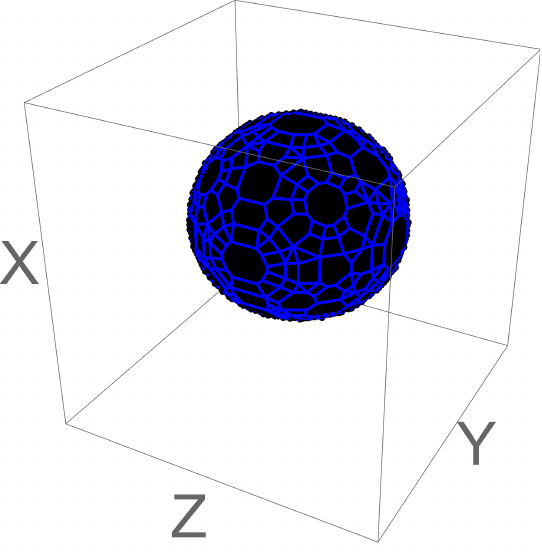}
    \caption{$N_{\rm pix}=64$}
  \end{subfigure}
  \hspace{0.00\textwidth}
  \begin{subfigure}[t]{0.3\textwidth}
    \centering
    \includegraphics[width=0.6\textwidth]{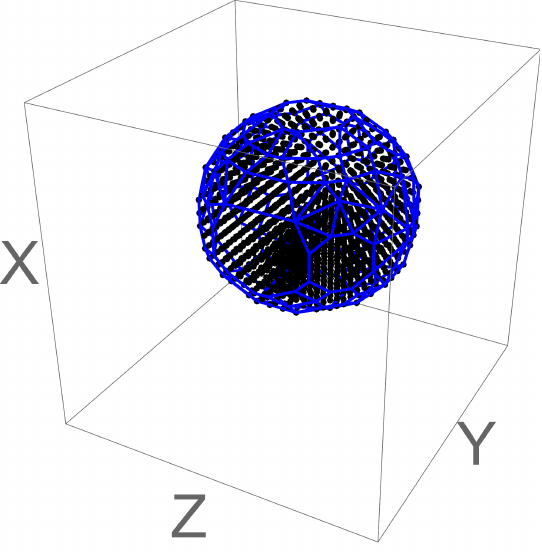}
    \caption{$N_{\rm pix}=24$}
  \end{subfigure}

  \caption{Hypershadows of the Myers-Perry black hole ($a=b=0.45$) rendered with different $N_{\rm pix}$ and viewpoints.  And the observer position is located at $\{65,\pi/2-1/20 \}$.}
  \label{MP_aa_hypershadow}
\end{figure}

In order to study the cohomogeneity-one Myers-Perry spacetime ($a=b$), the line element can be written as~\cite{myers2011myers}
\begin{equation}
\begin{aligned}
ds^2 &= -dt^2 + (x + a^2) \left( \frac{dx^2}{4\Delta} + d\theta^2 
+ \sin^2\theta\, d\phi^2 + \cos^2\theta\, d\psi^2 \right) \\
&\quad + \frac{\mu^2}{\rho^2} \left[ dt + a \left( \sin^2\theta\, d\phi + \cos^2\theta\, d\psi \right) \right]^2,
\end{aligned}
\end{equation}
Therefore, the black hole spin is constrained by \( |a| \leq \mu/2 \).
The horizon is located at $x_H= \frac{1}{2}(\mu^2-2a^2+\sqrt{\mu^4-4 a^2 \mu^2})$.\\
We present the coefficients of the locally treated basis as follows:
\begin{subequations}
\begin{align}
A^t &= \sqrt{\frac{a^2 + x}{(a^2 + x)^2 - x}+ 1}, \\
B^t &= \frac{a}{\sqrt{((a^2 + x)^2 - x)\left((a^2 + x)^2 + a^2\right)}}, \\
C^t &= \frac{a}{\sqrt{((a^2 + x)^2 - x)\left((a^2 + x)^2 + a^2\right)}}, \\
A^\phi &= \frac{1}{\sin\theta} \cdot \sqrt{\frac{a^2 \cos^2\theta + (a^2 + x)^2}{(a^2 + x)\left((a^2 + x)^2 + a^2\right)}}, \\
B^\phi &=- \frac{a^2 \sin\theta}{\sqrt{(a^2 + x)\left((a^2 + x)^2 + a^2\right)\left(a^2 \cos^2\theta + (a^2 + x)^2\right)}}, \\
A^\psi &= \frac{1}{\cos\theta} \cdot \sqrt{\frac{a^2 + x}{a^2 \cos^2\theta + (a^2 + x)^2}}, \\
A^x &= 2 \sqrt{\frac{(x + a^2)^2 - x}{x + a^2}}, \\
A^\theta &= \frac{1}{\sqrt{x + a^2}}.
\end{align}
\end{subequations}
By evolving the geodesic equations\cite{Frolov:2003en, Diemer:2014lba} derived from the Hamiltonian formalism under specified initial conditions, we reconstruct the 3D hypershadow using the backward ray-tracing method (see Fig.~\ref{MP_aa_hypershadow} ).
To probe the geometric structure of the hypershadow, we fix the observation angles \( \{\alpha, \beta, \zeta\} \) to zero and compute cross-sectional slices through its center along each coordinate plane, in order to examine and discuss their symmetries.

\begin{figure}[H]
  \centering

  \begin{subfigure}[t]{0.3\textwidth}
    \centering
    \includegraphics[width=0.6\linewidth]{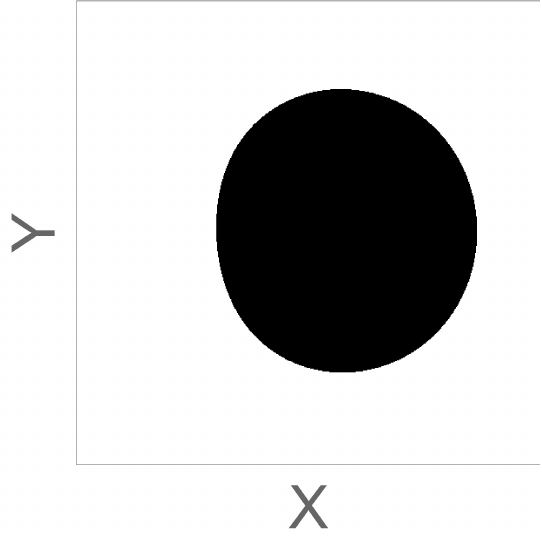}
    \caption{XY}
  \end{subfigure}
  \hspace{0\textwidth}
  \begin{subfigure}[t]{0.3\textwidth}
    \centering
    \includegraphics[width=0.6\linewidth]{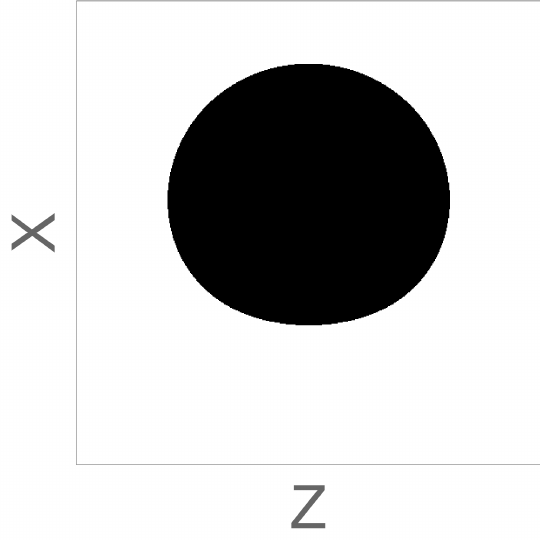}
    \caption{XZ}
  \end{subfigure}
  \hspace{0\textwidth}
  \begin{subfigure}[t]{0.3\textwidth}
    \centering
    \includegraphics[width=0.6\linewidth]{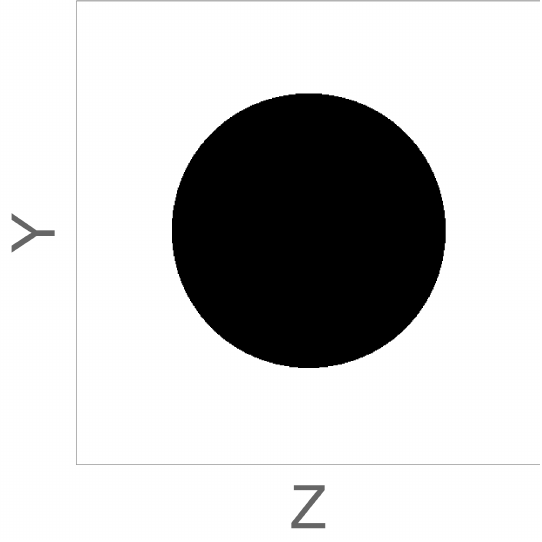}
    \caption{YZ}
  \end{subfigure}

  \vspace{0.5cm}

  \begin{subfigure}[t]{0.3\textwidth}
    \centering
    \includegraphics[width=0.6\linewidth]{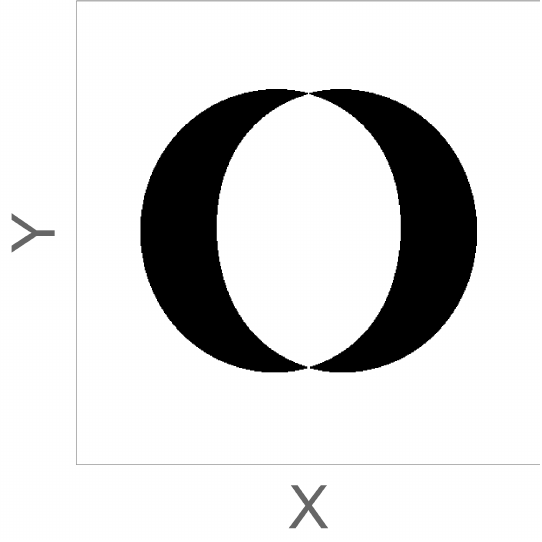}
    \caption{XY (H--$\Delta$)}
  \end{subfigure}
  \hspace{0.00\textwidth}
  \begin{subfigure}[t]{0.3\textwidth}
    \centering
    \includegraphics[width=0.6\linewidth]{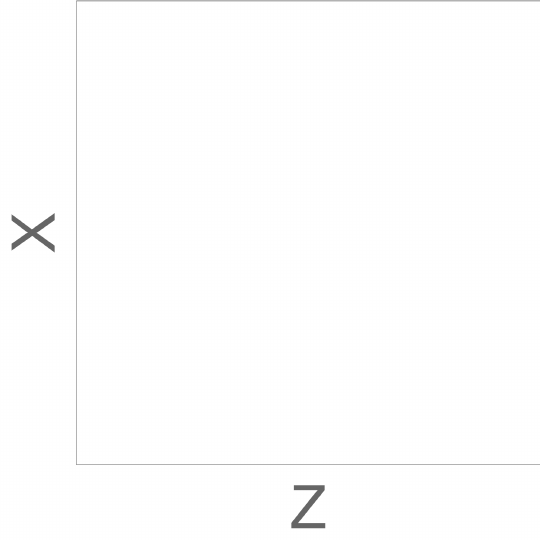}
    \caption{XZ (H--$\Delta$)}
  \end{subfigure}
  \hspace{0.00\textwidth}
  \begin{subfigure}[t]{0.3\textwidth}
    \centering
    \includegraphics[width=0.6\linewidth]{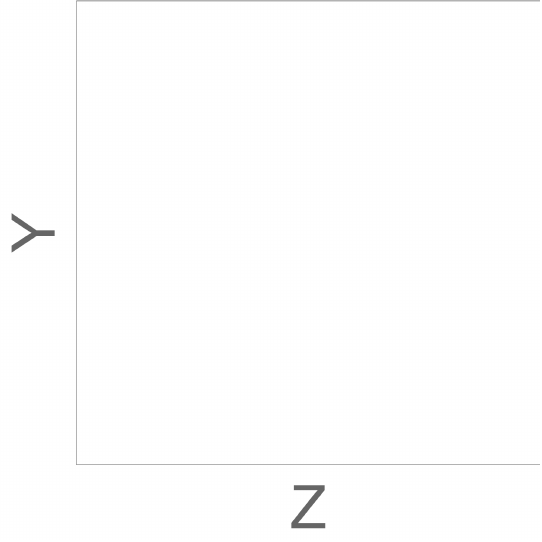}
    \caption{YZ (H--$\Delta$)}
  \end{subfigure}

  \vspace{0.5cm}

  \begin{subfigure}[t]{0.3\textwidth}
    \centering
    \includegraphics[width=0.6\linewidth]{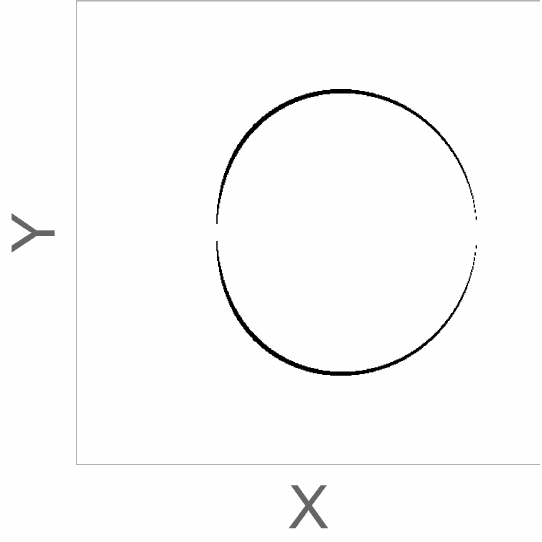}
    \caption{XY (V--$\Delta$)}
  \end{subfigure}
  \hspace{0\textwidth}
  \begin{subfigure}[t]{0.3\textwidth}
    \centering
    \includegraphics[width=0.6\linewidth]{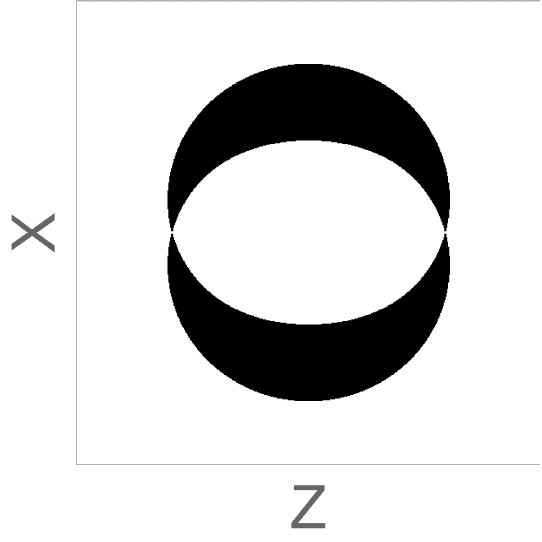}
    \caption{XZ (V--$\Delta$)}
  \end{subfigure}
  \hspace{0.00\textwidth}
  \begin{subfigure}[t]{0.3\textwidth}
    \centering
    \includegraphics[width=0.6\linewidth]{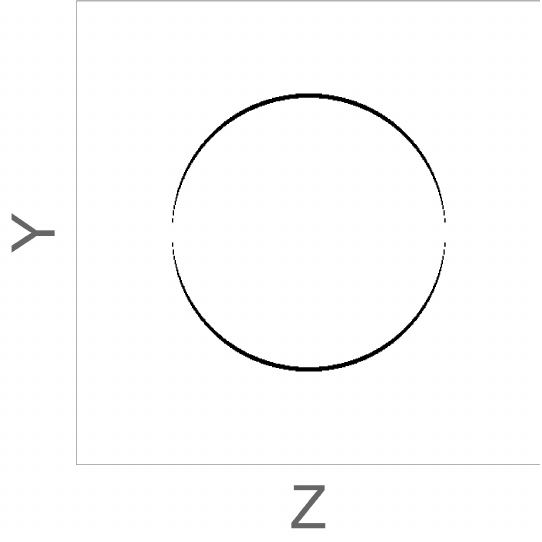}
    \caption{YZ (V--$\Delta$)}
  \end{subfigure}
\caption{
Top row: central slices of the hypershadow in the $XY$, $XZ$, and $YZ$ planes.
Middle row: H-$\Delta$ maps.
Bottom row: V-$\Delta$ maps.
Parameters: $a=b=0.45$, observer at $\{65,\pi/2-1/20\}$, and resolution $N_{\rm pix}=512$.}
\label{MP_aa_slices}
\end{figure}

From Fig.~\ref{MP_aa_slices},  we observe that the hypershadow exhibits mirror symmetry only with respect to the \(XY\) plane (i.e., under \( Z \mapsto -Z \)). This symmetry persists under different initial conditions.

In this paper \cite{Novo:2024wyn}, the authors proposed a simpler parametrization of the Myers-Perry (MP) hypershadow based on analytical construction. They concluded that the shape of the hypershadow is independent of the observer inclination angle \( \theta_{\rm o} \). Varying \( \theta_{\rm o} \) results only in a rigid rotation of the hypershadow in the image space, while the overall structure remains unchanged. Moreover, the geometry exhibits continuous rotational symmetry about the central axis.  When $\theta_{\rm o} = \pi/2$,  this central axis aligns with the $X$ axis.

To explore these symmetries numerically, we examine the difference maps under vertical reflection for various values of \( \theta_{\rm o} \).

\begin{figure}[H]
  \centering
  \begin{subfigure}[t]{0.3\textwidth}
    \centering
    \includegraphics[width=0.6\linewidth]{MP_aa_XY_UD_65.pdf}
    \caption{ $\theta_{\rm o}=\pi/2-1/20$}
  \end{subfigure}
  \hspace{0\textwidth}
  \begin{subfigure}[t]{0.3\textwidth}
    \centering
    \includegraphics[width=0.6\linewidth]{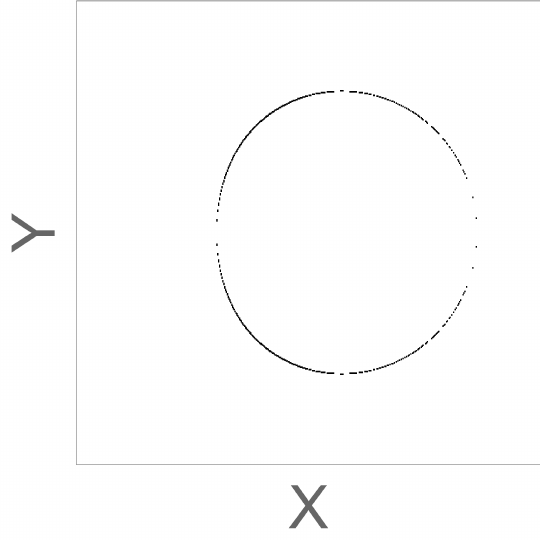}
    \caption{$\theta_{\rm o}=\pi/2-1/100$}
  \end{subfigure}
\caption{
$XY$ (H-$\Delta$) with different \( \theta_{\rm o} \).}
\label{MP_aa_diff}
\end{figure}

From Fig.~\ref{MP_aa_diff}, we find that as $\theta_{\rm o} \to \pi/2$ the difference between the original slice and its vertical reflection diminishes significantly. This suggests that the $XY$ central slice becomes symmetric under vertical reflection when $\theta_{\rm o} = \pi/2$. Consequently, this implies the existence of continuous rotational symmetry about the central axis. 
After confirming the approximate vertical symmetry near \( \theta_{\rm o} \approx \pi/2 \), we proceed to analyze the central slices in the \( XY \) plane for various observer inclination angles, in order to further illustrate the rotational behavior of the hypershadow.

\begin{figure}[H]
    \centering

    \begin{subfigure}[t]{0.18\textwidth}
        \centering
        \includegraphics[width=\linewidth]{MP_aa_XY_65_2pi.pdf}
        \caption*{$\theta_{\rm o}=\pi/2-1/20$}
    \end{subfigure}
    \begin{subfigure}[t]{0.18\textwidth}
        \centering
        \includegraphics[width=\linewidth]{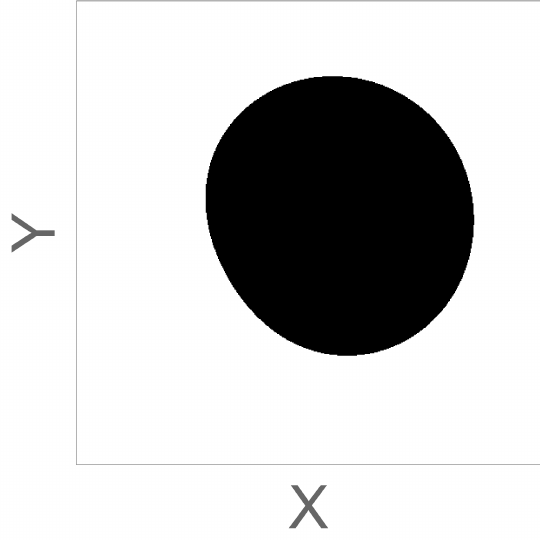}
        \caption*{$\theta_{\rm o}=\pi/3$}
    \end{subfigure}
    \begin{subfigure}[t]{0.18\textwidth}
        \centering
        \includegraphics[width=\linewidth]{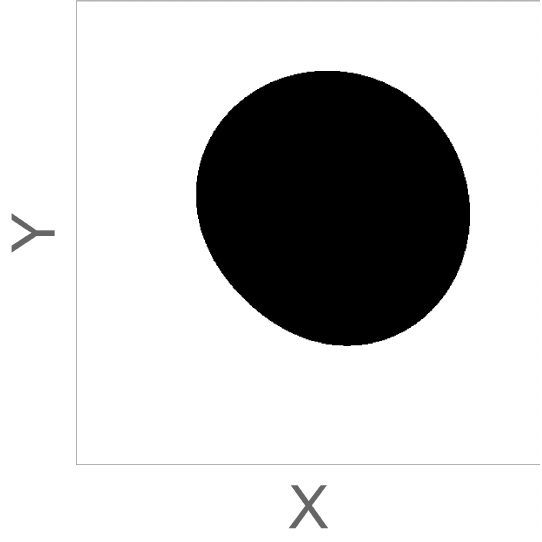}
        \caption*{$\theta_{\rm o}=\pi/4$}
    \end{subfigure}
    \begin{subfigure}[t]{0.18\textwidth}
        \centering
        \includegraphics[width=\linewidth]{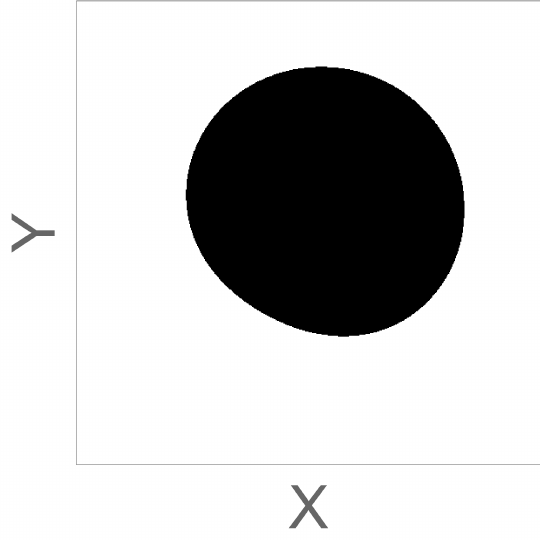}
        \caption*{$\theta_{\rm o}=\pi/6$}
    \end{subfigure}
    \begin{subfigure}[t]{0.18\textwidth}
        \centering
        \includegraphics[width=\linewidth]{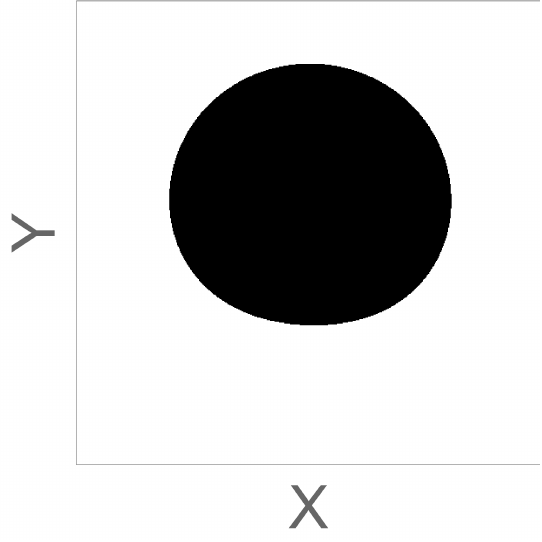}
        \caption*{$\theta_{\rm o}=1/20$}
    \end{subfigure}

    \caption{ $XY$ central slices with different $\theta_{\rm o}$.}
    \label{MP_aa_cent_diff}
\end{figure}

From Fig.~\ref{MP_aa_cent_diff}, we could get that when we decrease the value of $\theta_{\rm o}$ the central axis will counterclockwise rotates in the $XY$ center slice. As $\theta_{\rm o} \to 0$,  this central axis will align with the $Y$ axis.

In summary, the cohomogeneity-one Myers-Perry black hole (\(a = b\)) exhibits high degrees of symmetry in its hypershadow. The shape remains invariant under changes in the observer inclination angle \(\theta_{\rm o}\), resulting only in rigid rotations. Additionally, the hypershadow demonstrates mirror symmetry with respect to the \(XY\) plane and continuous rotational symmetry around the central axis, consistent with the enhanced Killing symmetries of the metric.

\subsubsection{Rotation in a single plane \texorpdfstring{$a=0$}{a=0}}
In addition to the cohomogeneity-one case, we further investigate the single-rotation Myers-Perry black hole by setting \( a = 0 \). This configuration is physically simpler, yet it lacks the additional Killing vectors that exist when \( a = b \). As a result, it is difficult to derive an analytical shadow boundary. To overcome this limitation, we apply our novel numerical approach to compute the 3D hypershadow in this case. This method not only fills the analytical gap, but also allows a detailed study of the shape and symmetry of the hypershadow under single plane rotation.

The line element for the single-rotation Myers-Perry black hole is given explicitly as follows\cite{Emparan:2008eg}:

\begin{equation}
\begin{aligned}
ds^2 = & -dt^2 + x \sin^2\theta\, d\phi^2 + (x + b^2)\cos^2\theta\, d\psi^2 \\
& + \frac{\mu^2}{\rho^2}\left(dt + b \cos^2\theta\, d\psi \right)^2 
+ \frac{\rho^2}{4\Delta} dx^2 + \rho^2 d\theta^2,
\end{aligned}
\end{equation}
with 
\begin{equation}
\rho^2 = x  + b^2 \sin^2\theta, 
\quad 
\Delta = x (x + b^2) - \mu^2 x.
\end{equation}
Therefore, the black hole spin is constrained by \( |b| \leq \mu \).
The horizon is located at $x_H= \mu^2 -b^2$.

\begin{figure}[H]
  \centering
  \begin{subfigure}[t]{0.3\textwidth}
    \centering
    \includegraphics[width=0.6\textwidth]{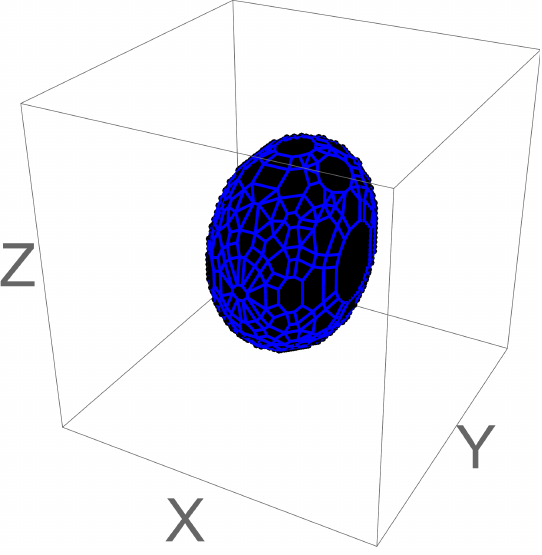}
    \caption{$N_{pix}=64$}
  \end{subfigure}
  \hspace{0.00\textwidth}
  \begin{subfigure}[t]{0.3\textwidth}
    \centering
    \includegraphics[width=0.6\textwidth]{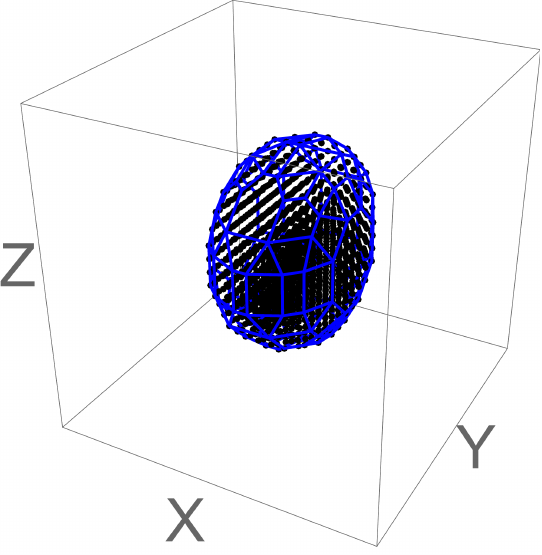}
    \caption{$N_{pix}=24$}
  \end{subfigure}

  \vspace{0.5cm}

  \begin{subfigure}[t]{0.3\textwidth}
    \centering
    \includegraphics[width=0.6\textwidth]{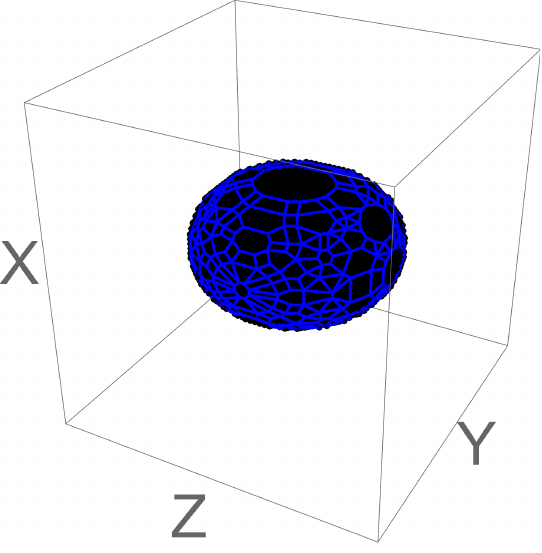}
    \caption{$N_{pix}=64$}
  \end{subfigure}
  \hspace{0.00\textwidth}
  \begin{subfigure}[t]{0.3\textwidth}
    \centering
    \includegraphics[width=0.6\textwidth]{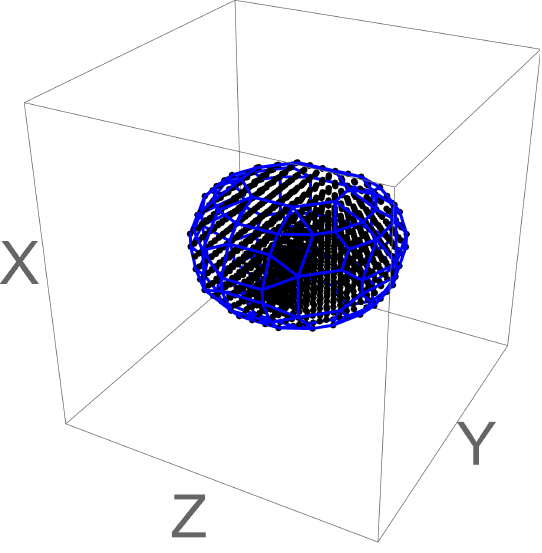}
    \caption{$N_{pix}=24$}
  \end{subfigure}
  \caption{Hypershadows of the Myers-Perry black hole ($a=0, b=0.95$) rendered with different $N_{\rm pix}$ and viewpoints.  And the observer position (the initial condition) is located at $\{ 65, \pi/2 - 1/20 \}$.}
  \label{MP_0b_hypershadow}
\end{figure}

At this point, the locally treated basic $\{\hat{e}_{(t)}, \hat{e}_{(\phi)} \}$ can be expressed in a simplified form, similar to that used in the Kerr black hole case: 
\begin{subequations}
\begin{align}
\hat{e}_{(t)} &= A^t \partial_t  + C^t \partial_\psi , \\
\hat{e}_{(\phi)} &= A^\phi \partial_\phi.
\end{align}
\end{subequations}
The coefficient $A^\phi$ is set as in Eq.~\eqref{A_psi}:
\begin{equation}
    A^\phi = \frac{1}{\sqrt{g_{\phi\phi}}} ,
\end{equation}
and then we could get
\begin{equation}
    A^t=\sqrt{\frac{g_{\psi\psi}}{g_{t\psi}^2-g_{tt} g_{\psi\psi}}}, \quad
    C^t=-\frac{g_{t\psi}}{g_{\psi\psi}} \sqrt{\frac{g_{\psi\psi}}{g_{t\psi}^2-g_{tt} g_{\psi\psi}}}.
\end{equation}
With the linear momentum $\mathbf{p}$ decomposition Eq.~\eqref{p_decomposition},  the conserved quantity of motion $\{\Phi, E\}$ can be written in a simplified form:
\begin{subequations}
\begin{align}
    \Phi &= \mathbf{p} \, \sqrt{g_{\phi\phi}} \cos \alpha \sin \beta ,\\
    E &= \mathbf{p} \, (\frac{1+C^t \sqrt{g_{\psi\psi}} \cos\alpha \cos\beta \sin\zeta}{A^t}).
\end{align}
\end{subequations}
Using the Hamilton equations\cite{Diemer:2014lba,Kagramanova:2012hw, Bugden:2018uya} together with the initial conditions, we employ the backward ray-tracing method to obtain the 3D hypershadow (see Fig.~\ref{MP_0b_hypershadow}).

To further investigate the geometric structure of the hypershadow, we fix the observation angles \( \{\alpha, \beta, \zeta\} \) to zero and compute cross-sectional slices through its center along each coordinate plane, in order to examine and discuss their symmetries.

\begin{figure}[H]
  \centering

  \begin{subfigure}[t]{0.3\textwidth}
    \centering
    \includegraphics[width=0.6\linewidth]{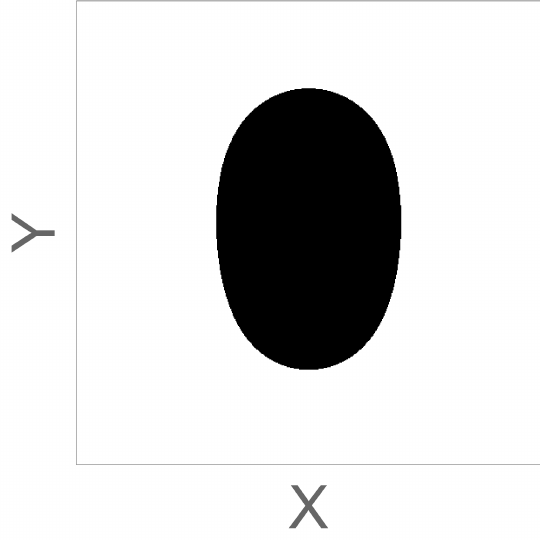}
    \caption{XY}
  \end{subfigure}
  \hspace{0.00\textwidth}
  \begin{subfigure}[t]{0.3\textwidth}
    \centering
    \includegraphics[width=0.6\linewidth]{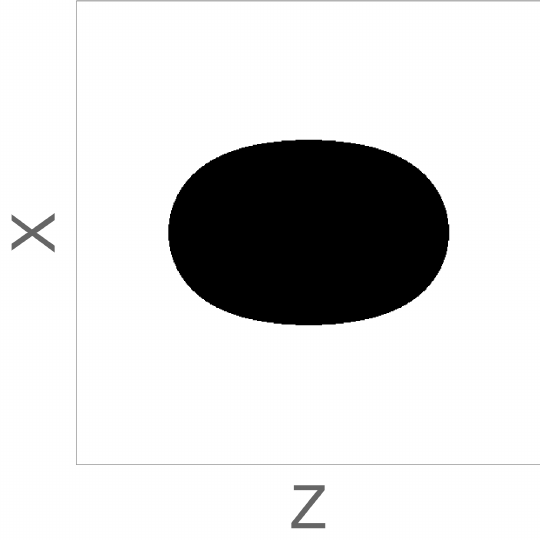}
    \caption{XZ}
  \end{subfigure}
  \hspace{0.00\textwidth}
  \begin{subfigure}[t]{0.3\textwidth}
    \centering
    \includegraphics[width=0.6\linewidth]{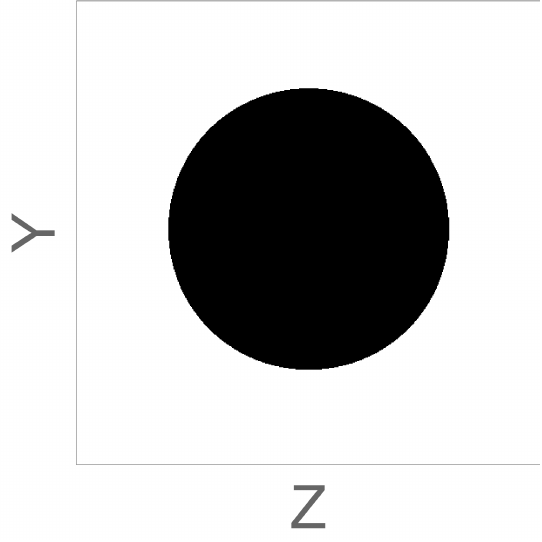}
    \caption{YZ}
  \end{subfigure}

  \vspace{0.5cm}

  \begin{subfigure}[t]{0.3\textwidth}
    \centering
    \includegraphics[width=0.6\linewidth]{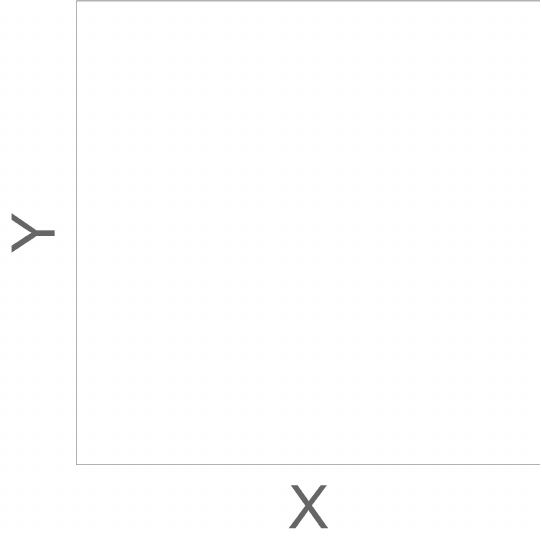}
    \caption{XY (H--$\Delta$)}
  \end{subfigure}
  \hspace{0.00\textwidth}
  \begin{subfigure}[t]{0.3\textwidth}
    \centering
    \includegraphics[width=0.6\linewidth]{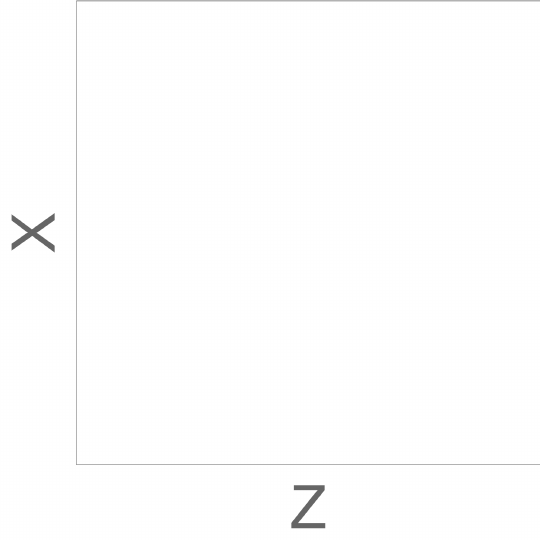}
    \caption{XZ (H--$\Delta$)}
  \end{subfigure}
  \hspace{0.00\textwidth}
  \begin{subfigure}[t]{0.3\textwidth}
    \centering
    \includegraphics[width=0.6\linewidth]{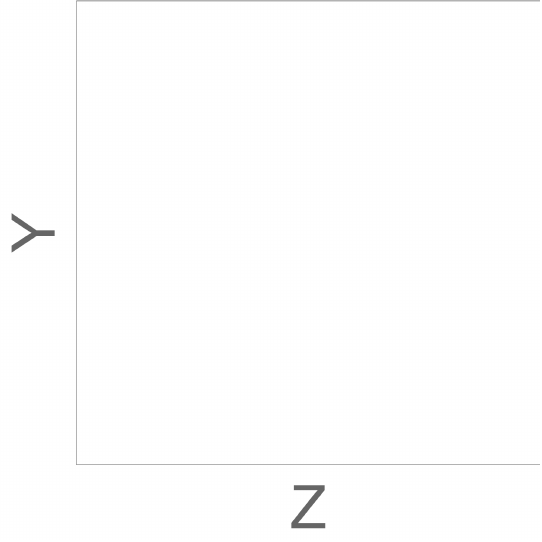}
    \caption{YZ (H--$\Delta$)}
  \end{subfigure}

  \vspace{0.5cm}

  \begin{subfigure}[t]{0.3\textwidth}
    \centering
    \includegraphics[width=0.6\linewidth]{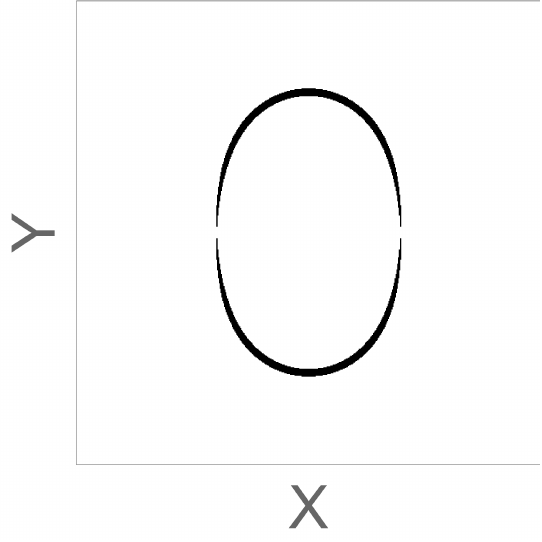}
    \caption{XY (V--$\Delta$)}
  \end{subfigure}
  \hspace{0.00\textwidth}
  \begin{subfigure}[t]{0.3\textwidth}
    \centering
    \includegraphics[width=0.6\linewidth]{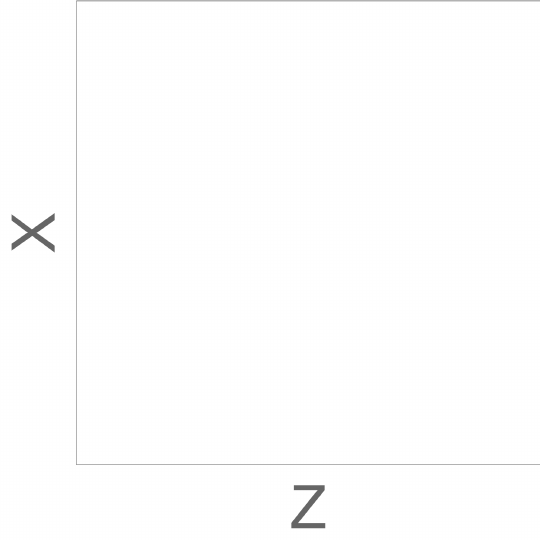}
    \caption{XZ (V--$\Delta$)}
  \end{subfigure}
  \hspace{0.00\textwidth}
  \begin{subfigure}[t]{0.3\textwidth}
    \centering
    \includegraphics[width=0.6\linewidth]{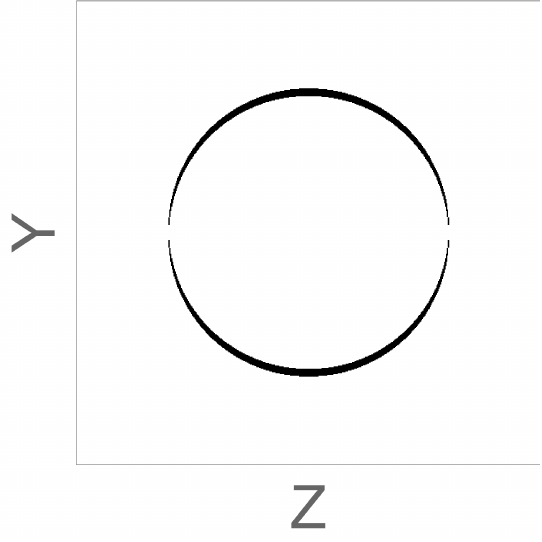}
    \caption{YZ (V--$\Delta$)}
  \end{subfigure}
\caption{
Top row: central slices of the hypershadow in the $XY$, $XZ$, and $YZ$ planes.
Middle row: H-$\Delta$ maps.
Bottom row: V-$\Delta$ maps.
Parameters: singly rotating case ($a=0$, $b=0.95$), observer at $\{65,\pi/2-1/20\}$, and resolution $N_{\rm pix}=512$.
}
\label{MP_0b_slices}
\end{figure}

From Fig.~\ref{MP_0b_slices}, we observe that the hypershadow exhibits mirror symmetry with respect to both the \(XY\) plane (i.e., under \(Z \mapsto -Z\)) and the \(YZ\) plane (i.e., under \(X \mapsto -X\)). These two mirror symmetries presist under different initial conditions.

Building on the discussion of continuous symmetry in the Myers-Perry black hole with equal angular momenta, we proceed to plot the difference maps under vertical reflection in the \( XY \) plane for \( \theta_{\rm o} = \pi/2 - 1/20 \) and \( \theta_{\rm o} = \pi/2 - 1/100\), in order to examine whether the single-rotation Myers-Perry black hole also exhibits continuous symmetry.

\begin{figure}[H]
  \centering
  \begin{subfigure}[t]{0.3\textwidth}
    \centering
    \includegraphics[width=0.8\textwidth]{MP_0b_XY_UD_65.pdf}
    \caption{$\theta_{\rm o}=\pi/2-1/20$}
  \end{subfigure}
  \hspace{0\textwidth}
  \begin{subfigure}[t]{0.3\textwidth}
    \centering
    \includegraphics[width=0.8\textwidth]{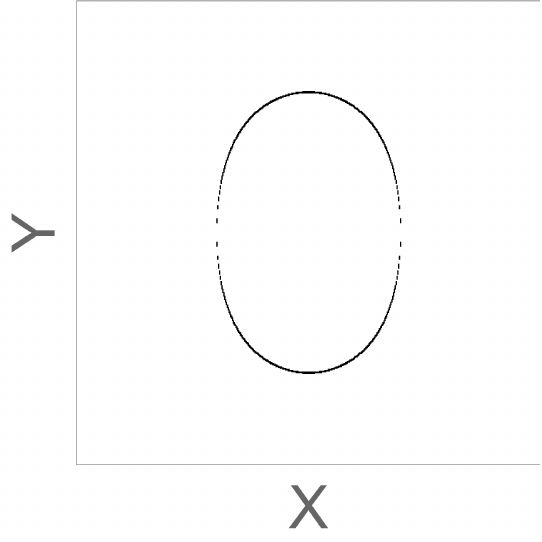}
    \caption{$\theta_{\rm o}=\pi/2-1/100$}
  \end{subfigure}
  \caption{$XY$ (H-$\Delta$) with different \( \theta_{\rm o} \).}
  \label{MP_0b_2D_diff}
\end{figure}

From Fig.~\ref{MP_0b_2D_diff}, we observe that as \( \theta_{\rm o} \to \pi/2 \) the difference between the original slice and its vertical reflection becomes negligible. This behavior closely resembles the \( a = b \) case, indicating that as \( \theta_{\rm o} \to \pi/2  \) the hypershadow approaches a configuration exhibiting continuous rotational symmetry about the \( X \) axis.

We next proceed to analyze the central slices in the \( XY \) plane for various observation angles, in order to further illustrate the deformation of the hypershadow.

\begin{figure}[H]
    \centering

    \begin{subfigure}[t]{0.18\textwidth}
        \centering
        \includegraphics[width=\linewidth]{MP_0b_XY_65_2pi.pdf}
    \end{subfigure}
    \begin{subfigure}[t]{0.18\textwidth}
        \centering
        \includegraphics[width=\linewidth]{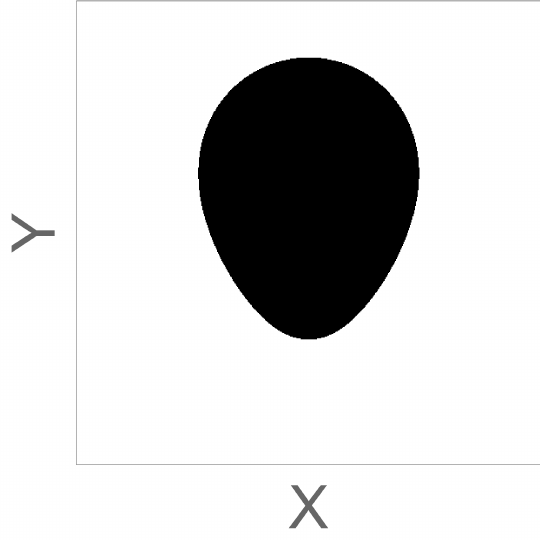}
    \end{subfigure}
    \begin{subfigure}[t]{0.18\textwidth}
        \centering
        \includegraphics[width=\linewidth]{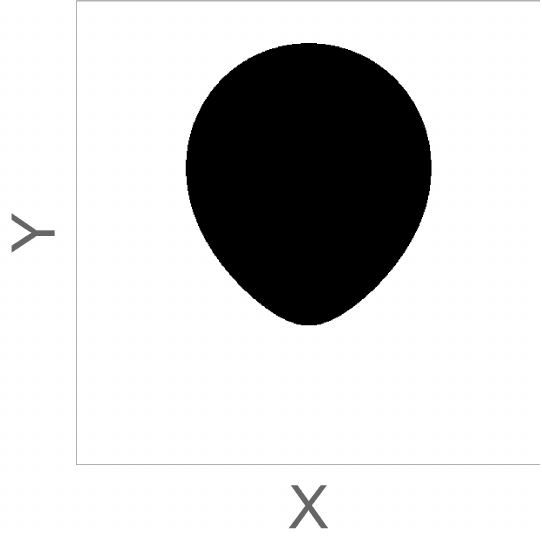}
    \end{subfigure}
    \begin{subfigure}[t]{0.18\textwidth}
        \centering
        \includegraphics[width=\linewidth]{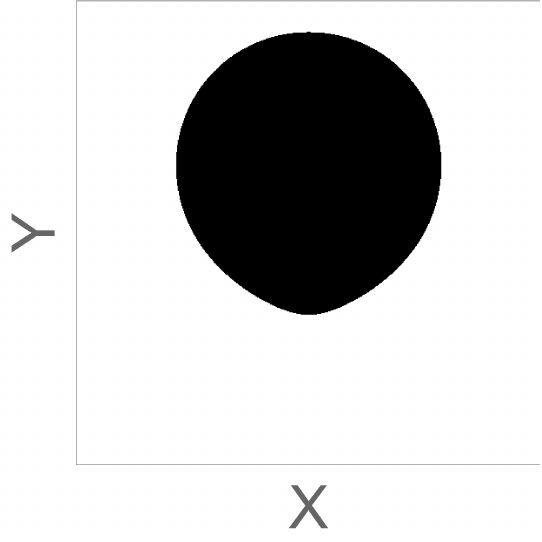}
    \end{subfigure}
    \begin{subfigure}[t]{0.18\textwidth}
        \centering
        \includegraphics[width=\linewidth]{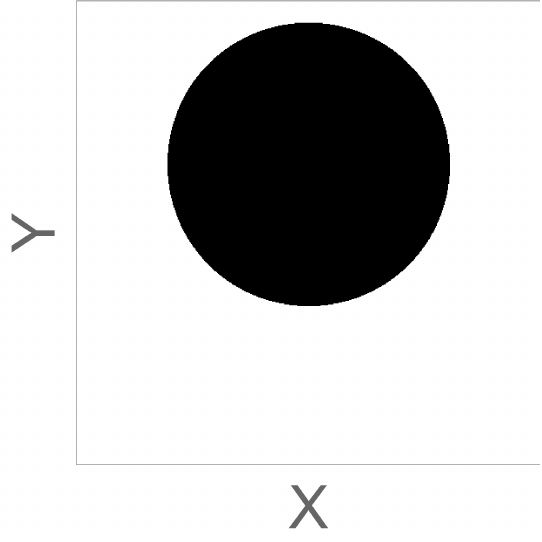}
    \end{subfigure}
    
    \vspace{0.5cm}

   \begin{subfigure}[t]{0.18\textwidth}
        \centering
        \includegraphics[width=\linewidth]{MP_0b_XZ_65_2pi.pdf}
    \end{subfigure}
    \begin{subfigure}[t]{0.18\textwidth}
        \centering
        \includegraphics[width=\linewidth]{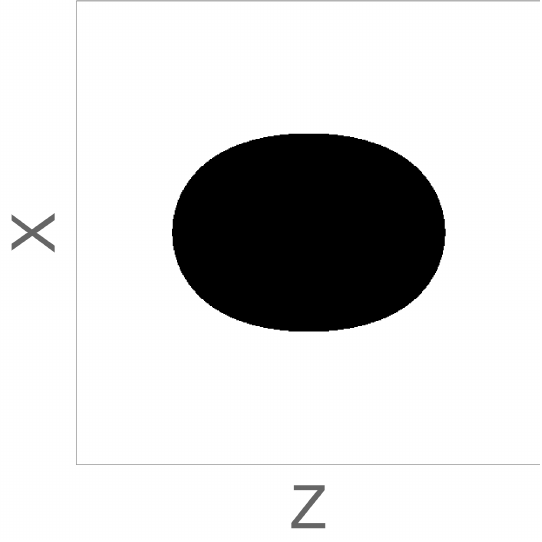}
    \end{subfigure}
    \begin{subfigure}[t]{0.18\textwidth}
        \centering
        \includegraphics[width=\linewidth]{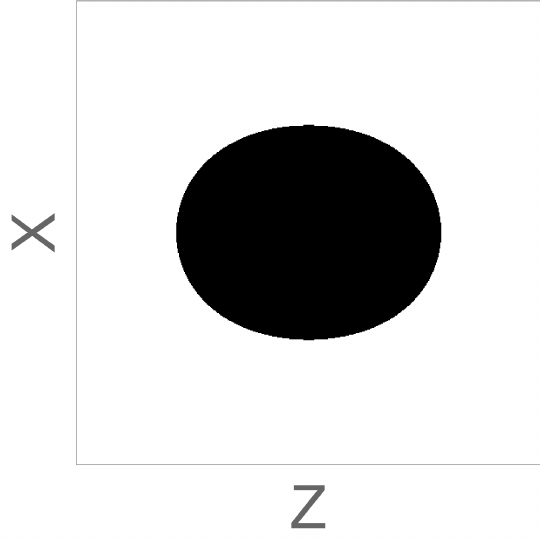}
    \end{subfigure}
    \begin{subfigure}[t]{0.18\textwidth}
        \centering
        \includegraphics[width=\linewidth]{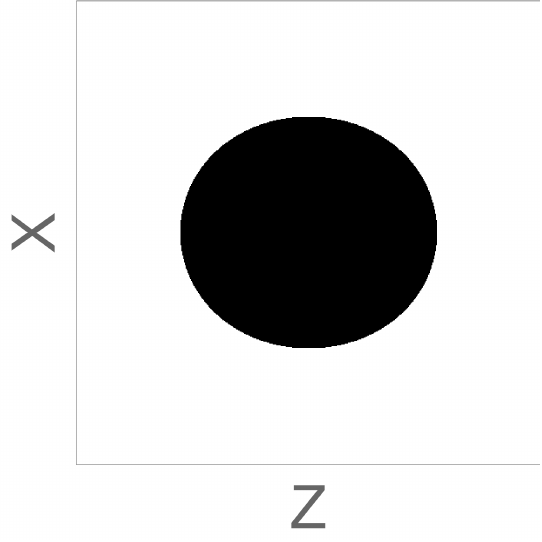}
    \end{subfigure}
    \begin{subfigure}[t]{0.18\textwidth}
        \centering
        \includegraphics[width=\linewidth]{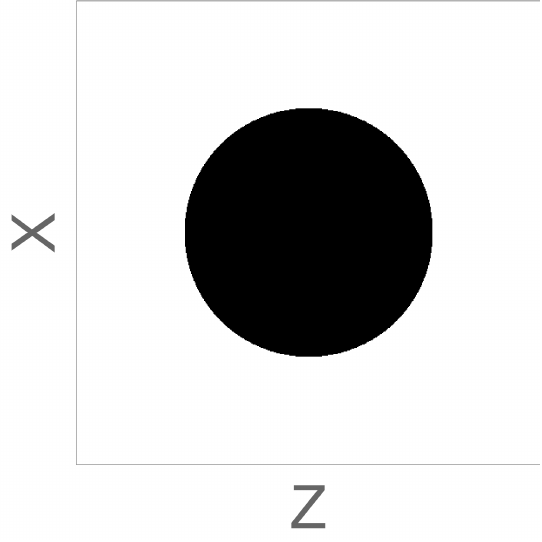}
    \end{subfigure}

    \vspace{0.5cm}

   \begin{subfigure}[t]{0.18\textwidth}
        \centering
        \includegraphics[width=\linewidth]{MP_0b_YZ_65_2pi.pdf}
        \caption*{$\theta_{\rm o} = \pi/2-1/20$}
    \end{subfigure}
    \begin{subfigure}[t]{0.18\textwidth}
        \centering
        \includegraphics[width=\linewidth]{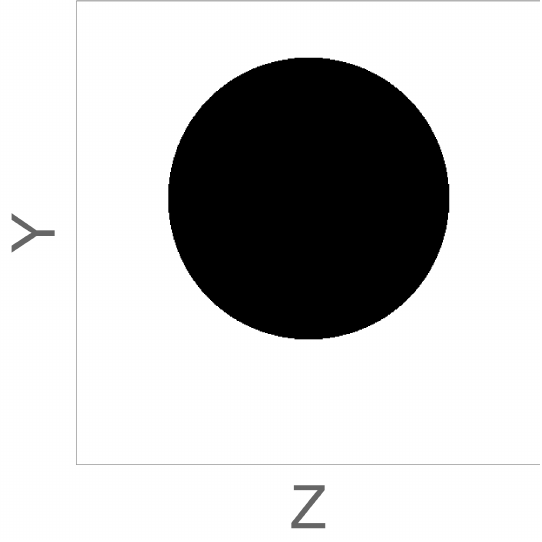}
        \caption*{$\theta_{\rm o} = \pi/3$}
    \end{subfigure}
    \begin{subfigure}[t]{0.18\textwidth}
        \centering
        \includegraphics[width=\linewidth]{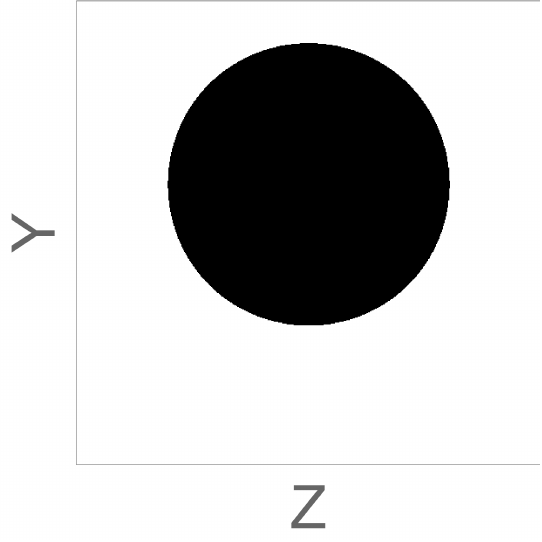}
        \caption*{$\theta_{\rm o} = \pi/4$}
    \end{subfigure}
    \begin{subfigure}[t]{0.18\textwidth}
        \centering
        \includegraphics[width=\linewidth]{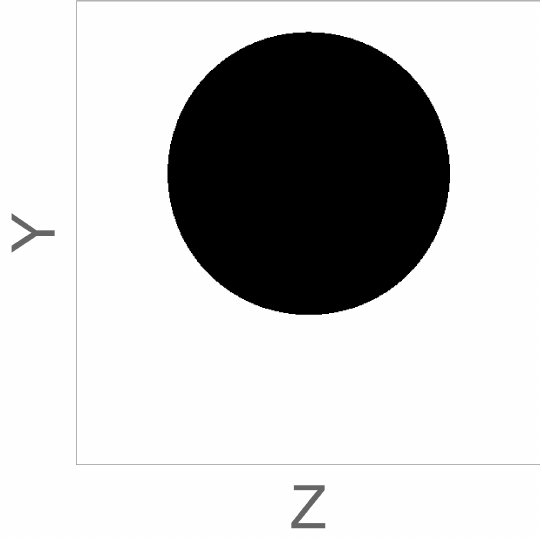}
        \caption*{$\theta_{\rm o} = \pi/6$}
    \end{subfigure}
    \begin{subfigure}[t]{0.18\textwidth}
        \centering
        \includegraphics[width=\linewidth]{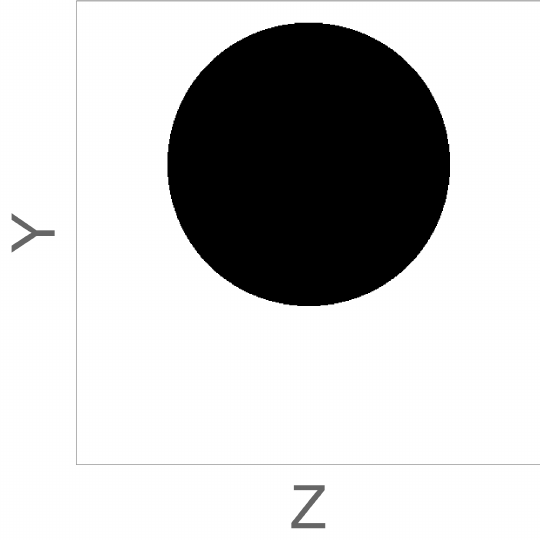}
        \caption*{$\theta_{\rm o} = 1/20$}
    \end{subfigure}
    \caption{Central slices for different values of \( \theta_{\rm o} \). The first row shows the central slices in the \( XY \) plane, the second row in the \( XZ \) plane, and the third row in the \( YZ \) plane.}
    \label{MP_0b_cent_diff}
\end{figure}

\begin{figure}[H]
    \centering

    \begin{subfigure}[t]{0.18\textwidth}
        \centering
        \includegraphics[width=\linewidth]{MP_0b_01_65_2pi.pdf}
    \end{subfigure}
    \begin{subfigure}[t]{0.18\textwidth}
        \centering
        \includegraphics[width=\linewidth]{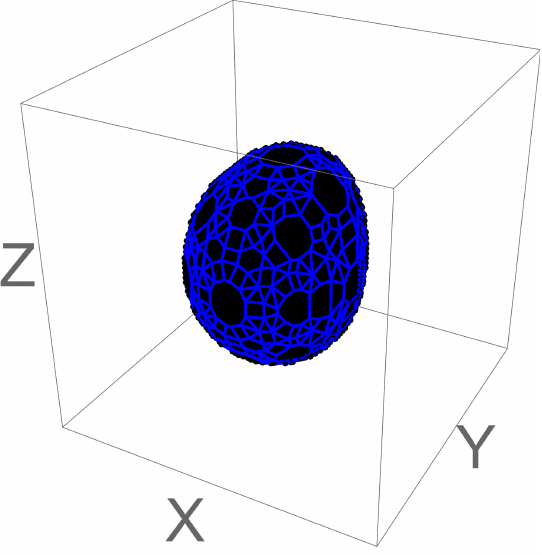}
    \end{subfigure}
    \begin{subfigure}[t]{0.18\textwidth}
        \centering
        \includegraphics[width=\linewidth]{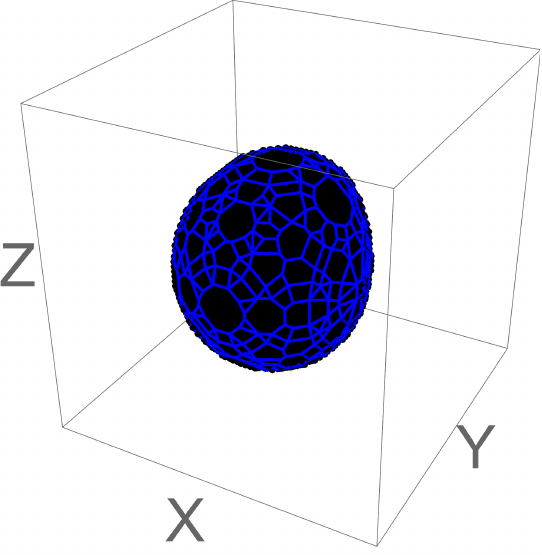}
    \end{subfigure}
    \begin{subfigure}[t]{0.18\textwidth}
        \centering
        \includegraphics[width=\linewidth]{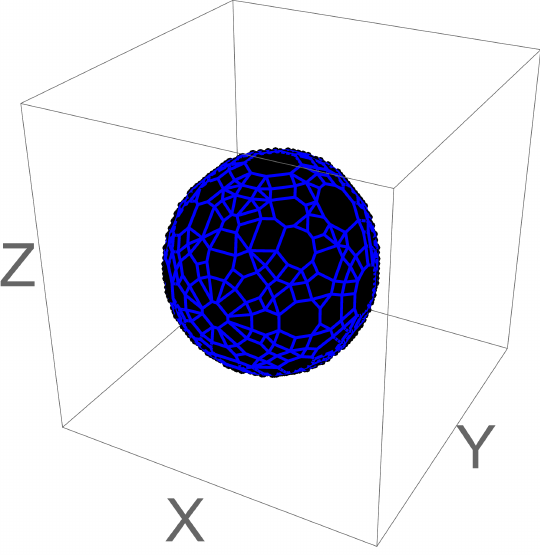}
    \end{subfigure}
    \begin{subfigure}[t]{0.18\textwidth}
        \centering
        \includegraphics[width=\linewidth]{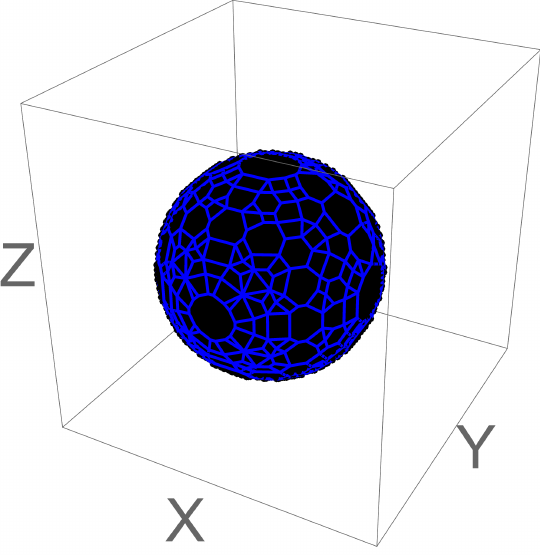}
    \end{subfigure}
    
    \vspace{0.5cm}

   \begin{subfigure}[t]{0.18\textwidth}
        \centering
        \includegraphics[width=\linewidth]{MP_0b_03_65_2pi.pdf}
        \caption*{$\theta_{\rm o}=\pi/2-1/20$}
    \end{subfigure}
    \begin{subfigure}[t]{0.18\textwidth}
        \centering
        \includegraphics[width=\linewidth]{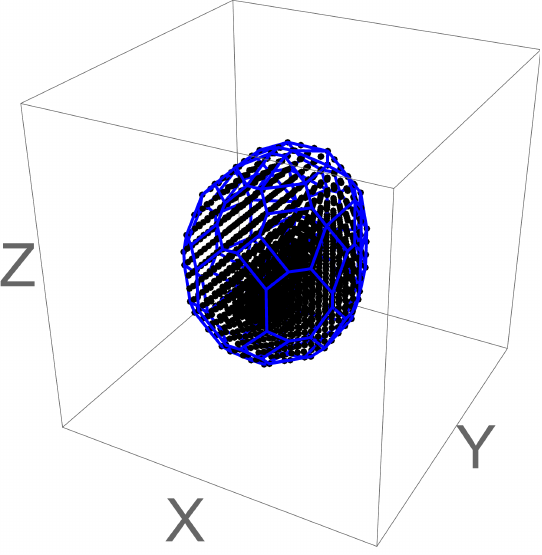}
        \caption*{$\theta_{\rm o}=\pi/3$}
    \end{subfigure}
    \begin{subfigure}[t]{0.18\textwidth}
        \centering
        \includegraphics[width=\linewidth]{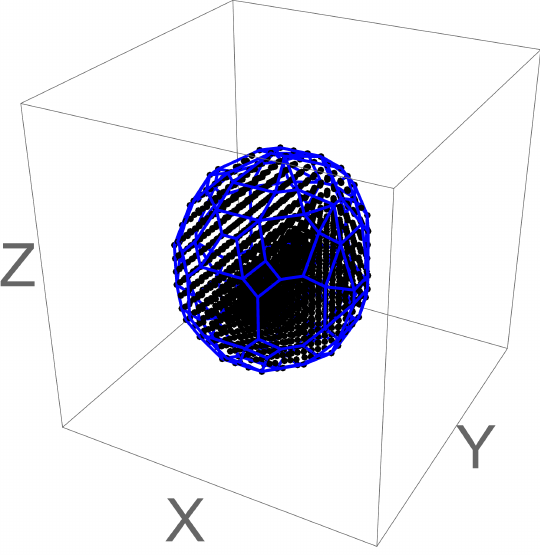}
        \caption*{$\theta_{\rm o}=\pi/4$}
    \end{subfigure}
    \begin{subfigure}[t]{0.18\textwidth}
        \centering
        \includegraphics[width=\linewidth]{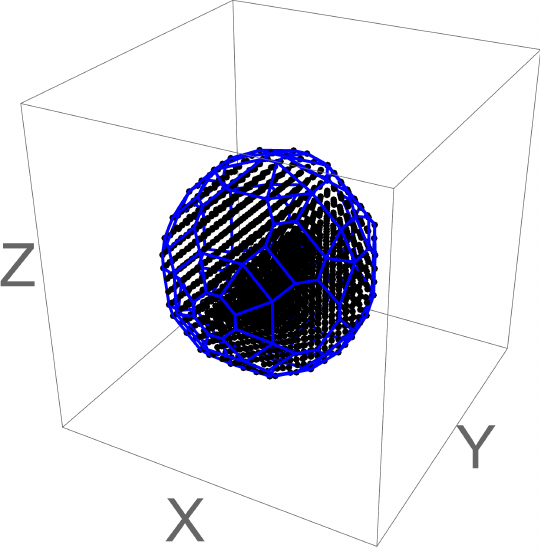}
        \caption*{$\theta_{\rm o}=\pi/6$}
    \end{subfigure}
    \begin{subfigure}[t]{0.18\textwidth}
        \centering
        \includegraphics[width=\linewidth]{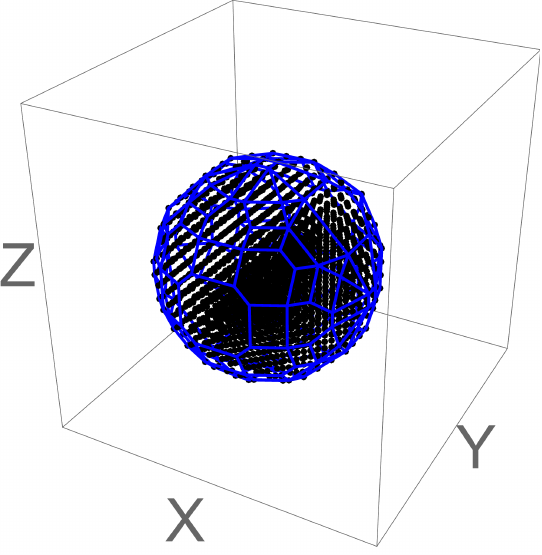}
        \caption*{$\theta_{\rm o}=1/20$}
   \end{subfigure}
    \caption{ Hypershadows of the Myers-Perry black hole ($a=0, b=0.95$) rendered with different $N_{\rm pix}$ and observer position. The resolutions is set to $N_{\rm pix}=64$ in the first row and $N_{\rm pix}=24$ in the second row.}
    \label{MP_0b_3D_diff}
\end{figure}

As \( \theta_{\rm o} \) decreases toward zero, the central slice in the \( XY \) plane undergoes a noticeable deformation. From Fig.~\ref{MP_0b_cent_diff}, we observe that the hypershadow of the Myers-Perry black hole gradually approaches a nearly circular shape, closely resembling that of the Tangherlini black hole. However, its center simultaneously shifts away from the image center.
These findings are further supported by the full three-dimensional hypershadow renderings presented in Fig.~\ref{MP_0b_3D_diff}.

In comparison, the hypershadow of the single-rotation Myers-Perry black hole (\(a = 0)\) exhibits a distinct behavior. While it retains mirror symmetry with respect to both the \(XY\)  and \(YZ\) planes, its shape gradually transitions toward a more spherical structure as \(\theta_{\rm o}\) decreases. 
Unlike the cohomogeneity-one case, where the hypershadow undergoes rigid rotation under changes in the observer inclination angle, the single-rotation configuration results in nontrivial shape deformation. 
These results highlight the richer geometric response in less symmetric spacetimes and demonstrate the flexibility of our numerical framework in capturing such variations.

\section{EFFECTS OF OBSERVER POSITION ON THE BLACK HOLE SHADOW}\label{effects}

In this section, we study how the observer position shapes the black hole shadow (hypershadow).
It is well known that increasing the observer radial coordinate $r_o$ (denoted $x_{\rm o} = r_o^2$ in five dimensions) decreases the apparent size of the shadow. 
The influence of the observer inclination angle $\theta_{\rm obs}$ has been analyzed extensively in Refs.~\cite{Vazquez:2003zm, Papnoi:2014aaa, Hu:2020usx}. They  work with angular (celestial) coordinates or impact-parameter mappings in more direct, conventional ways using a projection different from that adopted in our work. 
By contrast, in the equirectangular projection used in this paper, it has received little attention, with only a few studies\cite{Novo:2024wyn} addressing it. 
Here we present a detailed analysis of this effect within the equirectangular projection.

\begin{figure}[H]
  \centering
  \begin{subfigure}[t]{0.3\textwidth}
    \centering
    \includegraphics[width=0.6\textwidth]{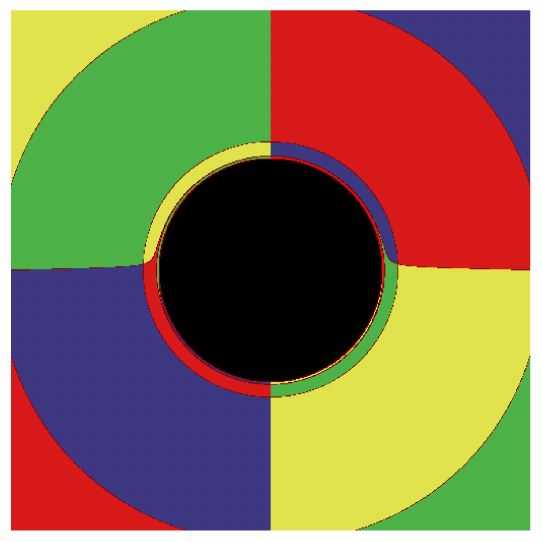}
    \caption{"SCH", $\theta_{\rm o}=\pi/2-1/20$}
  \end{subfigure}
  \hspace{0.00\textwidth}
  \begin{subfigure}[t]{0.3\textwidth}
    \centering
    \includegraphics[width=0.6\textwidth]{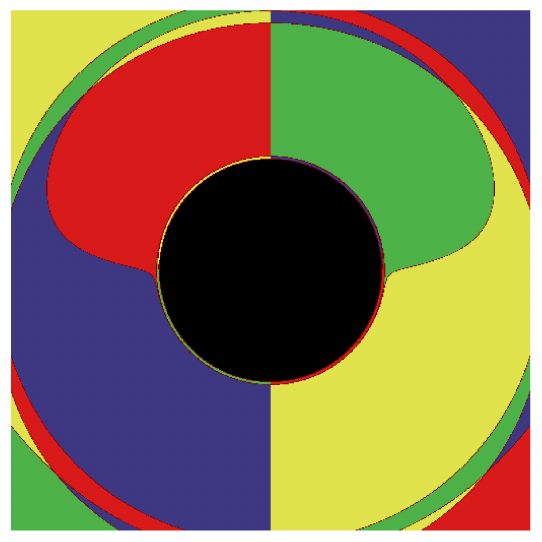}
    \caption{"SCH", $\theta_{\rm o}=1/20$}
  \end{subfigure}

  \vspace{0.5cm}

  \begin{subfigure}[t]{0.3\textwidth}
    \centering
    \includegraphics[width=0.6\textwidth]{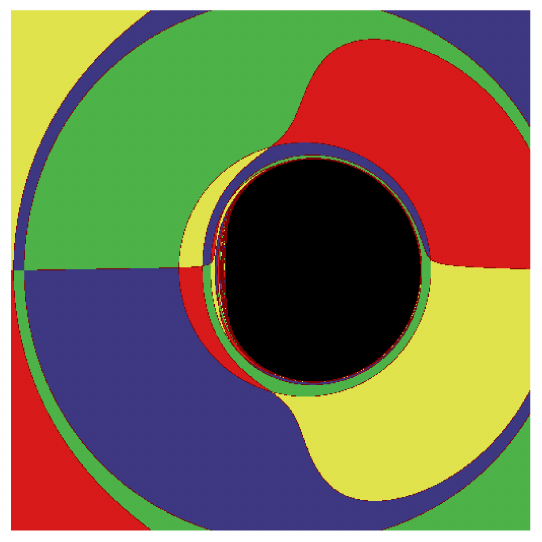}
    \caption{"KERR", $\theta_{\rm o}=\pi/2-1/20$}
  \end{subfigure}
  \hspace{0.00\textwidth}
  \begin{subfigure}[t]{0.3\textwidth}
    \centering
    \includegraphics[width=0.6\textwidth]{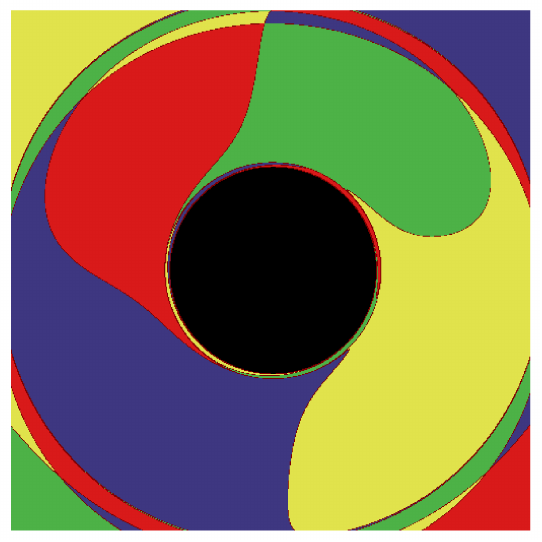}
    \caption{"KERR", $\theta_{\rm o}=1/20$}
  \end{subfigure}

  \caption{The shadows of the Schwarzschild and Kerr black holes with different observer inclination angle \( \theta_{\rm o} \) }
  \label{fig:SchKerrShadow}
\end{figure}

To set the stage, we briefly review the $\theta_{\rm o}$ dependence in the four-dimensional Schwarzschild and Kerr cases; see Fig.~\ref{fig:SchKerrShadow}.
When $\theta_{\rm o}\to 0$, the shadows become circular. 
In the Kerr case, frame dragging near the boundary distorts the photon trajectories, while the Schwarzschild case retains symmetric deflection without dragging.
These color-coded panels conveniently illustrate how light rays are gravitationally deflected and either are captured by the black hole or escape to infinity, thereby outlining the shadow boundary.

In higher dimensions such a visualization of individual light-ray trajectories is no longer feasible.
Instead, the impact of $\theta_{\rm o}$ can be inferred from two image-space diagnostics of the hypershadow: 
(i) \emph{deformation} (changes in shape) and (ii) \emph{displacement} (a global offset of the image). 
The corresponding five-dimensional results are shown in Figs.~\ref{MP_aa_cent_diff},  \ref{MP_0b_cent_diff}, and\ref{MP_0b_3D_diff}. 
In what follows we discuss the geometric origin of these features and then quantify the variations with observer inclination.

The physical origin can be traced to the two orthogonal rotation planes of the five-dimensional Myers-Perry geometry:
\[
d\Omega_3^2 = d\theta^2 + \sin^2\theta\, d\phi^2 + \cos^2\theta\, d\psi^2 ,
\]
where the $a$ spin acts in the $\phi$ plane (weighted by $\sin^2\theta$) and the $b$ spin in the $\psi$ plane (weighted by $\cos^2\theta$).
Here, $\theta$ is the $S^3$ angle controlling the local coupling of each spin, whereas $\theta_{\rm o}$ selects the mix sampled along the line of sight. 
In what follows, we focus on two representative spin configurations that already exhibit sharply different responses.

\noindent\emph{(i) Cohomogeneity-one ($a=b$).} 
The two planes contribute symmetrically; frame-dragging effects combine isotropically in the image space, so varying $\theta_{\rm o}$ produces a rigid rotation without displacement or distortion.

\noindent\emph{(ii) Single spin ($a=0$).}
In the singly rotating case ($a=0$), the two effects separate cleanly at the end points of $\theta_{\rm o}$.
When the observer is aligned with the rotation axis ($\theta_{\rm o}=0$), frame dragging acts almost uniformly in the local mapping of impact parameters, producing a global offset while the hypershadow boundary remains nearly circular.
By contrast, when the observer lies in the rotational equator ($\theta_{\rm o}=\pi/2$), the global offset vanishes and the boundary exhibits the strongest differential distortion.
For intermediate angles $0<\theta_{\rm o}<\pi/2$, both contributions are present: The centroid (visual-center) shift gradually decreases with $\theta_{\rm o}$, whereas the boundary distortion increases, yielding a smooth trade-off between translation and deformation.
Taken together, the $a=b$ case highlights the robustness of rotational symmetry, whereas the $a=0$ case shows that $\theta_{\rm o}$ is a sensitive probe of rotational anisotropy in higher dimensions.

We now quantify the impact of the observer inclination angle by scanning $\theta_{\rm o}$ jointly with the spin parameters $(a,b)$ and tracking two observables: (i) a distortion parameter $\delta_s$ that characterizes the hypershadow shape, and (ii) a displacement parameter $\eta$ that measures its global offset.

\paragraph{Cohomogeneity-one case ($a=b$) vs.\ Schwarzschild-Tangherlini.}
We first compare the voxel counts of the cohomogeneity-one Myers-Perry hypershadow with those of the Schwarzschild-Tangherlini (ST) black hole at the same $x_{\rm o}$ and resolution. 
The results show that all voxels of the MP hypershadow are included within the ST reference, with no additional points protruding beyond it. 
Accordingly, the total number of captured voxels in the MP case is smaller than in the ST case, indicating that rotation reduces the overall size of the hypershadow without producing any extra boundary features.

\paragraph{Single-spin case ($a=0$) vs.\ Schwarzschild-Tangherlini.}
For the singly rotating  Myers-Perry black hole, the hypershadow exhibits mirror symmetry with respect to both the $XY$ plane ($Z\mapsto -Z$) and the $YZ$ plane ($X\mapsto -X$); see Fig.~\ref{MP_0b_slices}. 
Because of this symmetry, the visual center can be determined simply as the midpoint between the top and bottom voxel along the $Y$ axis. 
We define the displacement $\eta$ as the distance of this midpoint from the origin. 
After shifting the ST reference by $\eta$ in the $Y$ direction, it still fully covers the MP hypershadow. 

Taken together, the results indicate that black hole spin primarily shrinks the hypershadow and produces a global offset, without generating new boundary features, consistent with earlier observations~\cite{Papnoi:2014aaa}. 
To probe subtle deviations from spherical symmetry in the hypershadow, we introduce a distortion parameter $\delta_s$. 
This definition, appropriate for the 3D setting, differs from the 2D shadow distortion employed in Ref.~\cite{Papnoi:2014aaa}.

We evaluate the total number of captured voxels at fixed radial coordinate $x_{\rm o}$ and resolution $N_{\rm pix}^3$.
Let $N_{\rm cap}^{\rm ST}$ and $N_{\rm cap}^{\rm MP}$ denote the captured-voxel counts for the
ST and MP cases, respectively.
We measure the deficit relative to the ST baseline by
\[
N_{\rm loss} \;:=\; N_{\rm cap}^{\rm ST} - N_{\rm cap}^{\rm MP}.
\]
The distortion parameter is then defined as the normalized deficit
\[
\delta_s \;=\; 100 \times \frac{N_{\rm loss}}{N_{\rm cap}^{\rm ST}},
\]
which quantifies the degree to which the hypershadow deviates in size from the ST reference. 
By construction, $\delta_s=0$ corresponds to equal size, while larger $\delta_s$ indicates a greater reduction of the captured region. Note that the distortion parameter $\delta_s$ is expressed as a percentage.

\begin{table}[H]
    \caption{Captured voxel statistics for the single-rotation Myers-Perry black hole.}
	\label{results}
	\centering
    \begin{tabular}{| c | c || c | c | c | c | c |}
        \hline
       \makecell{Observer \\ Inclination Angle}  & Voxel statistics & $b=0.95$ & $b = 0.8$ & $b = 0.6$ & $b=0.4$ & $b=0.2$\\ \hline

    \multirow{3}{*}{\makecell{$\pi/2-1/20$}}&  $N_{\rm cap}^{\rm MP}$ & 23832 & 27992 & 31404 & 33500 & 34748 \\
         \cline{2-7}
        
     & $N_{\rm loss}$ & 11280 & 7120 & 3708 & 1612 & 364
     \\  \cline{2-7}
     &  $\delta_s$  & 32.1 & 20.3 & 10.6 & 4.5 & 1.03
     \\\hline

    \multirow{3}{*}{\makecell{$\pi/3$}}&  $N_{\rm cap}^{\rm MP}$ & 26118 & 29778 & 32332 & 33928 & 34860 \\
         \cline{2-7}
        
     & $N_{\rm loss}$ & 8444 & 5324 & 2780 & 1184 & 252
     \\  \cline{2-7}
     & $\delta_s$  & 24 & 15.1 & 7.9  & 3.37 & 0.7
     \\\hline

    \multirow{3}{*}{\makecell{$\pi/4$}}&  $N_{\rm cap}^{\rm MP}$ & 29464 & 31516 & 33272 & 34328 & 34904 \\
         \cline{2-7}
        
     & $N_{\rm loss}$ & 5468 & 3596 & 1840 & 784 & 208
     \\  \cline{2-7}
     & $\delta_s$  & 16 & 10.2 & 5.2  & 2.2 & 0.59
     \\\hline

    \multirow{3}{*}{\makecell{$\pi/6$}}&  $N_{\rm cap}^{\rm MP}$ & 32288 & 33200 & 34196 & 34712 & 35000 \\
         \cline{2-7}
        
     & $N_{\rm loss}$ & 2824 & 1792 & 916  & 400 & 112
     \\  \cline{2-7}
     & $\delta_s$  & 8.04 & 5.1 & 2.61 & 1.14 & 0.31
     \\\hline
 
    \end{tabular}

\end{table}

At the working resolution $64^3$, the ST baseline yields
$N_{\rm cap}^{\rm ST}=35112$ captured voxels.
For the singly rotating Myers-Perry case ($a=0$), Table.~\ref{results}
summarizes, for each observer inclination and spin $b$, the captured-voxel counts
 $N_{\rm cap}^{\rm MP}$, the deficit $N_{\rm loss}$, and the distortion parameter $\delta_s $.
Two clear trends emerge: (i) At fixed $\theta_{\rm o}$, $\delta_s $ decreases
monotonically as $b$ is reduced ;
(ii) at fixed $b$, $\delta_s $ grows monotonically as the observer inclination angle increases. Moreover, increasing $b$ enhances the angular contrast: For larger $b$ the distortion parameter is well described by a $\sin^2\,\theta_{\rm o}$ dependence, whereas for smaller $b$ the proportionality is less evident.
\begin{table}[H]
    \caption{Myers-Perry captured-voxel statistics, $\theta_{\rm o}=\pi/2-1/20$: $a=0$  vs.\ $a=b$.}
	\label{results2}
	\centering
    \begin{tabular}{| c | c || c | c | c | c | c |}
        \hline
       \makecell{Case}  & Voxel statistics & $b=0.45$ & $b = 0.4$ & $b = 0.3$ & $b=0.2$ & $b=0.1$\\ \hline

    \multirow{3}{*}{\makecell{$a=0$ }}&$N_{\rm cap}^{\rm MP}$ & 23832 & 27992 & 31404 & 33500 & 34748 \\
         \cline{2-7}
        
     & $N_{\rm loss}$ & 11280 & 7120 & 3708 & 1612 & 364
     \\  \cline{2-7}
     & $\delta_s$ & 32.1 & 20.3 & 10.6 & 4.5 & 1.03
     \\\hline

    \multirow{3}{*}{\makecell{$a=b$}}& $N_{\rm cap}^{\rm MP}$ & 26118 & 29778 & 32332 & 33928 & 34860 \\
         \cline{2-7}
        
     & $N_{\rm loss}$ & 8444 & 5324 & 2780 & 1184 & 252
     \\  \cline{2-7}
     & $\delta_s$ & 24 & 15.1 & 7.9  & 3.37 & 0.7
     \\\hline

\end{tabular} 
\end{table}

Table.~\ref{results2} reports, for the near-equatorial view ($\theta_{\rm o} \approx \pi/2$), the captured-voxel count $N_{\rm cap}^{\rm MP}$, the deficit $N_{\rm loss}$, and the distortion parameter $\delta_s$ for both the cohomogeneity--one case ($a=b$) and the single-rotation case ($a=0$). From these data we infer that, at fixed spin $b$, the distortion in the cohomogeneity-one case is always greater than or equal to that in the single-rotation case.

Having established the size trend, we now turn to the displacement.
Specifically, we evaluate the center shift (displacement) $\eta$ on the central $YZ$ slice ($X=0$) for the single-rotation configuration ($a=0$), using an in-slice resolution of $512^2$.

\begin{table}[H]
	\caption{Displacement of the hypershadow in the single-rotation case. }
	\label{displacement}
	\centering
		\begin{tabular}{| c || c | c | c | c | c |}
			\hline
			\makecell{Observer \\ Inclination Angle} & $b=0.95$ & $b=0.8$ & $b=0.6$ & $b=0.4$ & $b=0.2$ \\
			\hline
			$\pi/2-1/20$ & 4 & 3.5 & 2.5 & 2 & 1 \\
			\hline
			$\pi/3$ & 39 & 32.5 & 25 & 16.5 & 8 \\
			\hline
			$\pi/4$ & 55.5 & 46.5 & 35 & 23 & 12 \\
			\hline
			$\pi/6$ & 68 & 57 & 43 & 29 & 14 \\
			\hline
			$1/20$ & 78 & 66 & 49.5 & 33 & 16.5 \\
			\hline
	\end{tabular}
\end{table}
From Table~\ref{displacement}, for the singly rotating Myers-Perry black hole ($a=0$), we find that the hypershadow displacement $\eta$ increases as the observer inclination decreases at fixed rotation parameter $b$ and, at fixed $\theta_{\rm o}$, grows monotonically with $b$.

Relative to the Schwarzschild-Tangherlini baseline, spin reduces the hypershadow size: The distortion parameter $\delta_s$ increases with $b$ in both configurations and in the single-rotation case ($a=0$) also grows with the observer inclination. 
Displacement is specific to $a=0$: The center shift $\eta$ increases with $b$ and strengthens as $\theta_{\rm o}$ decreases, whereas for $a=b$ the hypershadow undergoes a rigid rotation with no displacement.

\section{CONCLUSION}
\label{sec5}
Extensive studies have clarified the optical appearance of four-dimensional black holes, culminating in direct observational confirmation by the Event Horizon Telescope (EHT). This naturally raises the question of how such features generalize to higher-dimensional settings.
In this work, we introduced a numerical approach based on backward ray tracing to compute the hypershadow, the three-dimensional generalization of the black hole shadow in five-dimensional spacetimes. Our method enables systematic control over observer parameters and facilitates the reconstruction of the full shadow as a three-dimensional volumetric object. The method allows systematic control of observer parameters and reconstructs the hypershadow as a volumetric object. 
For visualization, we combined discrete sampling with surface contouring and introduced reflection difference maps to quantify mirror symmetries.

Applying this framework to the Schwarzschild-Tangherlini and Myers-Perry solutions, we confirmed the exact spherical symmetry of the Schwarzschild-Tangherlini hypershadow and revealed the contrasting behaviors of Myers-Perry geometries. 
The cohomogeneity-one case (\(a=b\)) exhibits rigid rotational invariance, while the single-rotation case (\(a=0\)) shows clear size reduction, displacement, and symmetry breaking that grow with inclination and spin. 
To quantify these effects, we introduced two observables: a distortion parameter, measuring size deficit relative to the Tangherlini baseline, and a displacement parameter, characterizing global offsets.

Our findings demonstrate that hypershadows provide a powerful probe of the rotational structure and symmetry properties of higher-dimensional black holes. 
The numerical framework presented here establishes a robust foundation for future numerical explorations of more exotic objects, such as black rings and their prospective toroidal hypershadows.

\section*{ACKNOWLEDGMENTS}
The author gratefully acknowledges Prof. Carlos Herdeiro and Dr. Pedro Cunha for their guidance, helpful discussions, and valuable comments on the manuscript.
The author also gratefully acknowledges helpful discussions and support from Dr. João Novo, Dr. Ivo Sengo, and Dr. Zhen Zhong.

This work is supported by CIDMA under the FCT Multi-Annual Financing Program for R\&D Units (grants UID/4106/2025 and UID/PRR/4106/2025), through the Portuguese Foundation for Science and Technology (FCT -- Fundaç\~ao para a Ci\^encia e a Tecnologia), as well as the projects: Horizon Europe staff exchange (SE) programme HORIZON-MSCA2021-SE-01 Grant No.\ NewFunFiCO-101086251;  2022.04560.PTDC (\url{https://doi.org/10.54499/2022.04560.PTDC}) and 2024.05617.CERN (\url{https://doi.org/10.54499/2024.05617.CERN}).
This work is supported by the China Scholarship Council.

\bibliographystyle{ieeetr}  
\bibliography{references}
\end{document}